\documentstyle[prl,aps]{revtex} \def\narrowtext{} \tighten\twocolumn
\newcommand{\REF}{\mbox{\ref{eq:diagabrupt}--\ref{eq:vertabrupt}}}
\begin{document}
\draft
\title{
\begin{minipage}[t]{7.0in}
\scriptsize
\begin{quote}
\leftline{{\it Phys. Rev.} {\bf B}, in press (2003)}
\raggedleft{\rm OUTP-02-01S}\\
\raggedleft {\rm cond-mat/0206257}
\end{quote}
\end{minipage}
\medskip
\\Magnetic Domain Walls in Single-Phase and Phase-Separated
Double Exchange Systems}
\author{D. I. Golosov\thanks{E-mail: golosov@thphys.ox.ac.uk}}
\address{Theoretical Physics, Oxford University, 1 Keble Rd., Oxford
OX1 3NP, United Kingdom}  
\address{%
\begin{minipage}[t]{6.0in}
\begin{abstract}
We investigate the structure of magnetic domain walls in a classical
double exchange ferromagnet, evaluating domain wall energies
and charges. Three different cases are studied:
(i) a conventional smooth Bloch wall, (ii) an abrupt Ising-type wall,
which is shown to have lower energy at small values of carrier concentration,
and (iii) stripe wall, corresponding to the two ferromagnetic domains
being separated by a stripe of another, antiferromagnetic, phase.
General aspects of energy balance and geometry of phase-separated
states are discussed in this context.
It is speculated that domain walls of the latter type may be responsible
for the unusual transport properties of certain manganate films.
\typeout{polish abstract}
\end{abstract}
\end{minipage}}

\maketitle
\narrowtext
\newpage
\section{INTRODUCTION}
\label{sec:intro}

The unusual micromagnetic properties of colossal magnetoresistance (CMR) 
compounds are presently subject to intensive experimental
investigation\cite{deLozanne,Fukumura,Kwon,Welp,Mathur99,Li,Wu,Mathur01,Gupta,Soh2000,Miller}. 
In these studies, special attention is paid to the interplay between 
magnetic domain structure and transport properties of the system. Aside
from possible technological applications (associated with the large
low-field magnetoresistance\cite{Li}), the strong effect of magnetic
domain walls on conduction properties, as found in strained epitaxial
films of ${\rm La_{0.7}Ca_{0.3}MnO_3}$  (Refs. \cite{Mathur99,Li}),
 ${\rm Pr_{2/3}Sr_{1/3}MnO_3}$ (Ref. \cite{Li}), and ${\rm
La_{0.7}Sr_{0.3}MnO_3}$ (Refs. \cite{Li,Wu}), raises a genuine physical
problem. Indeed, given the 
relatively small expected value of the easy-axis magnetic anisotropy,
the usual Bloch (or N\'{e}el) domain wall would be rather smooth and broad. Thus, carrier
scattering off the Bloch walls could not appreciably affect  transport 
properties of the system. The measurement of magnetic domain walls 
contribution to the resistivity therefore leads to the 
conclusion\cite{Littlewood} that the domain walls
arising in the samples studied in Refs. \cite{Mathur99,Li,Wu} have an
unusual, 
non-Bloch 
structure. It has even been suggested \cite{Li} that the 
double exchange interaction, which is responsible for the ferromagnetism of 
doped manganese oxides, cannot possibly account for such
poorly-conducting magnetic domain walls. While the origins of this suggestion
may be traced to the widespread but ill-founded notion that the magnetic 
properties of  
double exchange systems can be adequately described by an effective 
Heisenberg model, the peculiar physics of domain walls in double exchange 
ferromagnets has not yet been addressed theoretically. 

In the present article, we consider the standard single-orbital double
exchange  
model with the following Hamiltonian:

\begin{eqnarray}
{\cal H}=&-&\frac{t}{2} \sum_{\langle i,j \rangle,\alpha} 
\left(c^\dagger_{i \alpha}c_{j \alpha} +c^\dagger_{j \alpha}c_{i
\alpha}\right) - 
 \frac{J_H}{2S} \sum_{i, \alpha, \beta} \vec{S}_i
\vec{\sigma}^{\alpha \beta} c^\dagger_{i\alpha} c_{i\beta}
+ \nonumber \\
&+&\frac{J}{S^2}\sum_{\langle i, j \rangle} \vec{S}_i \vec{S}_j -
\frac{K}{2S^2} \sum_i \left(S^z_i \right)^2\,.
\label{eq:Ham}
\end{eqnarray}
Here  $c_{j \alpha}$ (with $\alpha=\uparrow, \downarrow$) are the electron 
annihilation operators, and
the vector $\vec{\sigma}^{\alpha \beta}$ is composed of Pauli
matrices. $J_H$ is the strength of Hund's rule ferromagnetic coupling between
the spins of carriers and the core spins $\vec{S}_i$, 
which also interact with each other via the
direct antiferromagnetic Heisenberg exchange $J$. 
The core spins are assumed
to be classical ($S\gg 1$), and the easy-axis single-ion anisotropy
$K/S^2$ is included
in order to account for the finite Bloch wall energy.
The lattice is assumed
to be square, which is thought to be more appropriate than the 
three-dimensional cubic one for modelling the
thin films studied experimentally; the extension of our analysis to the
three-dimensional case is straightforward but cumbersome,
and is expected to yield similar conclusions.  
The electron spectrum
in the ferromagnetic state is given by\cite{axis}
$\epsilon^{\uparrow,\downarrow}_{\vec{k}}=\epsilon_{\vec{k}}\mp J_H/2$
with $\epsilon_{\vec{k}}=-t(\cos k_1+ \cos k_2)$.
We consider the
experimentally relevant {\it half-metallic} case, when owing to a
sufficiently large value of $J_H$, the carrier band in the
ferromagnetic phase is completely
spin-polarised. Thus the value of chemical potential, denoted
$\mu-J_H/2$, must lie below the bottom of the spin-up subband, $\mu <J_H
-2$.  We note that $t$ in Eq. (\ref{eq:Ham}) corresponds to $2t$ in a different
notation sometimes used elsewhere in the literature; it should also be pointed
out that below, 
the conduction electron (rather than hole) density is denoted by $x$.
Throughout the paper we use units in which hopping $t$
and the lattice spacing are equal to unity, and we consider the
zero-temperature ($T=0$) case. 

Below we consider domain walls of three different types, 
which are relevant for different values of parameters characterizing the
double exchange magnet at low temperatures. These are conventional Bloch
walls, abrupt (Ising) walls, and stripe walls, formed by a stripe of
antiferromagnetic phase inserted between the two ferromagnetic domains.

 We begin in Sect. \ref{sec:bloch} 
with the usual smooth
Bloch wall. The Bloch wall energy
depends on spin stiffness $D$ and anisotropy strength in a usual way
\cite{Volume8}, 
reflecting
the fact that the long-wavelength properties of double exchange ferromagnets 
are adequately captured within an effective Heisenberg description (cf. 
Ref. \cite{prl00}).
In double exchange systems, Bloch walls carry an electric charge, which
we also evaluate. Our results suggest that magnetic domain walls
arising in homogeneous (single-phase) double exchange ferromagnets at
the intermediate doping levels typically have Bloch structure,
and therefore cannot significantly affect the resistance of the sample.

Within the context of recent experiments, the possibility of sharp
changes in magnetisation direction within a domain wall is of
particular relevance. This scenario has been discussed for a long time
(see, {\it e.g.}, Ref. \cite{Zhang}) and it is  important to
consider it in some detail. Therefore in Sect. \ref{sec:abrupt} we treat the
extreme case of an abrupt
(Ising-type) domain wall, when the sign of magnetisation is reversed
over one lattice link.  A wall of this type, which in the $T=0$, 
$S \rightarrow \infty$ limit is impenetrable for carriers,
would  strongly affect the transport properties of the system.
The energy cost of an abrupt wall originates from
the underlying non-perturbative scattering problem for conduction
electrons. The corresponding physics is thus completely
non-Heisenberg. We derive expressions for energies and charges of
abrupt domain walls running in  two different directions (along 
a crystal axis and diagonally), and for all values of the Hund's rule
exchange constant, $J_H$. While for small values of carrier density,
$x\ll 1$,
the energy of an abrupt wall is lower than that of a Bloch wall (which
may be relevant for certain magnetic semiconductors), this does
not generally hold at the intermediate doping levels. In the latter
case, an abrupt domain wall is preferred only for very large values of
anisotropy $K \sim DS$, or for the case of very finely tuned parameter values,
providing for an almost exact balance between the ferro- and
antiferromagnetic tendencies of the system. It would
be unrealistic to expect that such a fine-tuning (within one per cent
in the values of $J$, $J_H$, and $x$ in a single-phase system) can be
achieved by different experimental groups in a reproducible way.  

In addition, it also turns out that these parameter values typically 
correspond to the system being unstable with respect to phase separation. As
explained in Sect. \ref{sec:stripe}, the latter phenomenon has a double effect:
(i) the carrier density within the bulk of the
ferromagnet is now determined by the condition that the thermodynamic
potentials of the two phases must be equal to each other; this condition
effectively pins the parameters of double exchange ferromagnet in the
region where the energy of an abrupt domain wall is relatively low.
(ii) Energy of an abrupt domain wall can be further lowered by
inserting a stripe of antiferromagnetic phase between the two
ferromagnetic domains. Since the two phases are characterized by
different values of charge
density, one cannot treat this situation properly without taking into
account the effects of Coulomb interaction. We use a somewhat
simplified treatment to estimate the energy and width of a stripe
domain wall. It turns out that within a certain range of parameter
values, the energy of a stripe wall can be lower than that of a Bloch
wall, so that magnetic domain walls in a phase-separated system are
actually of the stripe type. In particular, this situation is realized when
the antiferromagnetic phase occupies an appreciable area of the sample
(of the order of 15 \% of the net area, or possibly more), provided that
the easy-axis anisotropy constant $K$ is not too small. 
Due to insulating properties of the 
antiferromagnetic phase, carrier transport across the stripe wall is strongly
suppressed, leading to a substantial domain wall contribution to 
the sample resistance. On the other hand, ferromagnetic
area within a single magnetic domain remains well-connected, 
and phase separation is therefore not expected to
significantly affect the {\it intra-domain} metallic conductivity.
Analysis of the data of Refs. \cite{Mathur99,Li}
reveals a correlation between the film thickness, dielectric properties of the
substrate, and  the appearance of domain wall resistance, which seems
to agree with anticipated conditions for the stabilisation of the
stripe walls.

Details of calculations are relegated to the Appendices, which also
include a brief discussion of the three-dimensional case. 

The relevance of our findings in the context of recent experiments on
manganate films is further discussed in
Sect. \ref{sec:conclu}. We suggest that the domain walls observed
indirectly in the transport measurements of Refs. \cite{Mathur99,Li,Wu},
and directly in Ref. \cite{Mathur01}  are in fact
the stripe walls, introduced in Sect. \ref{sec:stripe}.

\section{BLOCH WALL}
\label{sec:bloch}

The structure of  domain walls in conventional Heisenberg ferromagnets
has been understood long ago\cite{Volume8}. These are smooth, long-wavelength
Bloch walls\cite{Neel}, and their surface tension (energy per unit length) 
$S_B$
and width $l_B$  are determined by the spin stiffness $D$
of the system:
\begin{equation}
S_B=2\sqrt{K\cdot (DS)}\,,\,\,\,\,\,\,l_B=\sqrt{DS/K}\,.
\label{eq:bloch}
\end{equation}
Since the unusual transport properties of the domain walls are found
only in certain strained films at a specific doping 
level\cite{Mathur99,Li,Wu}, we expect that in
most cases, domain walls in the CMR materials also have Bloch-like
structure. We will now study the relationship between the properties of
Bloch walls and the
parameters of our model Hamiltonian, Eq. (\ref{eq:Ham}).

The appropriate value of $D$ can be extracted from the
known  spin wave spectrum of a classical double exchange
ferromagnet\cite{Nagaev70} (see also Ref. \cite{prl00}):
\begin{equation}
\omega_{\vec{p}}=\frac{J_H}{2NS}\sum_{\vec{k}}n_{\vec{k}}
\frac{\epsilon_{\vec{k}}-\epsilon_{\vec{k}+\vec{p}}}{\epsilon^\uparrow_{\vec{k}}-\epsilon^\downarrow_{\vec{k}+\vec{p}}}+\frac{K}{S}+\frac{2J}{S}
\sum_{\alpha=1}^d(\cos k_\alpha -1)\,.
\label{eq:magnon}
\end{equation}
Here, $N$ is the number of lattice sites, and  $n_{\vec{p}}$ is the 
Fermi distribution
 function. Eq. (\ref{eq:magnon}) is valid for any dimensionality $d$
 and for an arbitrary electron dispersion law $\epsilon_{\vec{k}}$ (with
 $\epsilon_{\vec{k}}^{\uparrow,\downarrow}=\mp J_H/2 +\epsilon_{\vec{k}}$). 
 For the case  of the 2D tight-binding model (\ref{eq:Ham}), we obtain: 
\begin{eqnarray}
SD=&&-J+\left(\frac{|E|}{8}-\frac{1}{4J_H
N}\sum_{\vec{k}}n_{\vec{k}}v^2_{\vec{k}} 
\right)=-J-\frac{x}{4J_H}-
\nonumber \\
&&-\frac{\mu}{8\pi^2}\left(\mu-\frac{2+\mu^2}{J_H}\right)Y_1
+ \frac{1}{4\pi^2} \left(2-\frac{3\mu}{J_H} \right)Y_2\,.
\label{eq:stiffness}
\end{eqnarray}
Here and below, $Y_1$ and $Y_2$ denote the following complete elliptic
integrals:
\begin{equation}
Y_1={\cal K}\left(\sqrt{1-\frac{1}{4}[\mu(x)]^2}\right)\,,\,\,\,\,\,\,
Y_2={\cal E}\left(\sqrt{1-\frac{1}{4}[\mu(x)]^2}\right)\,,
\label{eq:elliptic}
\end{equation}
$\vec{v}_{\vec{k}}= \partial \epsilon_{\vec{k}}/\partial \vec{k}$ is
the electron velocity, and the kinetic energy of the band is given by
\begin{equation}
E\equiv \frac{1}{N} \sum_k n_{\vec{k}}
\epsilon_{\vec{k}}=\frac{\mu^2}{\pi^2}Y_1(x)- \frac{4}{\pi^2}Y_2(x).
\label{eq:bandenergy}
\end{equation}
Note that because
of the numerical prefactor entering Eq. (\ref{eq:stiffness}), the value of
$D$ is at least an order of magnitude smaller than that of the band
energy, $E$.

At low doping level, $x\ll 1 \stackrel{<}{\sim}J_H$,
Eq. (\ref{eq:stiffness}) 
yields
\begin{equation}
DS=-J+ \frac{1}{4}x-\frac{1}{8}\pi x^2- \frac{\pi x^2}{2 J_H}\,,
\label{eq:stifflowx}
\end{equation}
whereas at half-filling, $x=1$, we obtain
\begin{equation}
DS=- J -1/(4J_H)\,.
\label{eq:stiffhalf}
\end{equation}
The second term in Eq. (\ref{eq:stifflowx}), which is proportional to
 the
band energy ($E \approx -2x$ at low $x$) represents the leading-order
double-exchange (ferromagnetic) contribution. The last terms
in Eqs. (\ref{eq:stifflowx}--\ref{eq:stiffhalf}) indicate that the
effect of finite $J_H$ (as opposed to $J_H \rightarrow \infty$) is
similar to that of an increase in the value of direct superexchange,
 $J$. This conclusion is justified physically, since at  finite $J_H$
 an effective antiferromagnetic
 interaction arises due to virtual transitions between the two
 components of the spin-split band much like a usual superexchange,
 which is due to transitions between different bands. 
 Below we will
 see  how this qualitative analogy\cite{qualify} manifests itself
 in other properties of the system -- its validity is clearly not restricted 
to the spin stiffness evaluation. This in turn suggests that many of the 
 features of (more complicated) finite-$J_H$ systems can be modelled
 by treating the $J_H\rightarrow \infty$ case with an appropriately
 increased $J$.

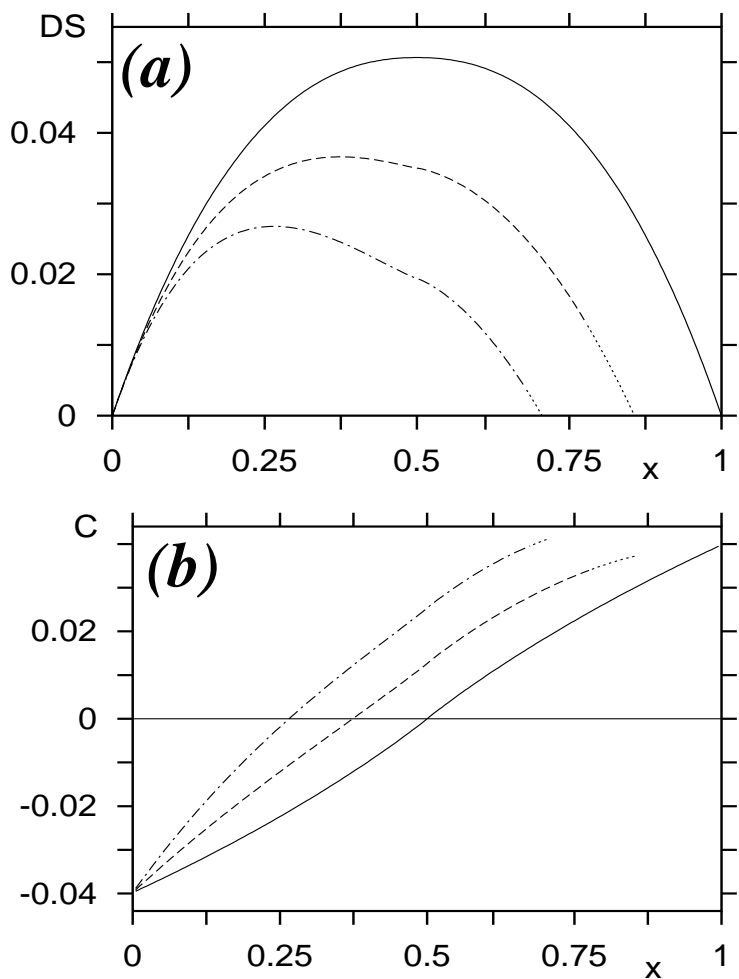
\begin{figure} 
\caption{Spin stiffness $DS$ {\it (a)} and the coefficient ${\cal C}$
[see Eq. (\ref{eq:blochcharge})] {\it (b)} vs.
electron density $x$ for $J=0$ and $J_H \rightarrow \infty$ (solid line),
$J_H=8$ (dashed line), and $J_H=4$ (dashed-dotted line). Dotted lines
correspond to a regime where the spin stiffness is still positive,
$D>0$, but the ferromagnetic phase is unstable with
respect to phase separation.}
\label{fig:bloch}
\end{figure}

The doping dependence of spin stiffness for three different values of
$J_H$ ($J_H\rightarrow\infty$,$J_H=8$,and $J_H=4$ for solid, dashed,
and dashed-dotted lines,  
respectively) and $J=0$ is shown in Fig. \ref{fig:bloch}
(a). For the case of finite $J_H$, the competition between effective
antiferromagnetism and double exchange-induced ferromagnetism, taking
place at sufficiently small $1-x$, is resolved via phase separation
\cite{Nagaevbook,dagottoreview,jap02,jap00}. This means that the homogeneous
ferromagnetic state becomes thermodynamically unstable as the electron
concentration   $x$ exceeds certain critical value. 
In Fig. \ref{fig:bloch} (a), the
values of $DS$ within the respective thermodynamically unstable regions
are plotted 
with dotted lines.  
When the
superexchange $J>0$ is present, this critical
value, which depends also on $J_H$, decreases further. In addition,
another region of phase-separation instabilities arises at low {\it
electron} densities\cite{dagottoreview,jap02}. 

Within a Bloch wall, misalignment of the neighbouring ionic spins leads to a
renormalisation of carrier hopping coefficient\cite{Anderson}.
Indeed, the Hamiltonian (\ref{eq:Ham}) can be re-written in terms of
new fermions $d_{i \uparrow}$ (and $d_{i\downarrow}$), whose spin is
aligned (antialigned) with the classical ionic spin $\vec{S}_i$ at the same 
site:
\begin{eqnarray}
&&{\cal H}=-\frac{1}{2} \sum_{\langle i,j \rangle,\alpha , \beta} 
\left(t_{ij}^{\alpha \beta} d^\dagger_{i \alpha}d_{j \beta} +
t_{ji}^{\alpha \beta} d^\dagger_{j \alpha}d_{i
\beta}\right) + \label{eq:Ham1} \\
&&+ \frac{J_H}{2} \sum_{i} \left( d^\dagger_{i\downarrow}
d_{i\downarrow} -d^\dagger_{i\uparrow}
d_{i\uparrow}\right)
+\frac{J}{S^2}\sum_{\langle i, j \rangle} \vec{S}_i \vec{S}_j -
\frac{K}{2S^2} \sum_i \left(S^z_i \right)^2\,.
\nonumber
\end{eqnarray}
Here, the matrix $t^{\alpha \beta}$ is given by 
\begin{eqnarray}
t_{ij}^{\alpha \beta}&=&\left( \begin{array}{cc} 
\tilde{C}_i \tilde{C}_j +{\rm e}^{{\rm i}(\phi_j
-\phi_i)} \tilde{S}_i \tilde{S}_j &
-{\rm e}^{-{\rm i} \phi_j} \tilde{C}_i \tilde{S}_j
+{\rm e}^{-{\rm i} \phi_i} \tilde{S}_i \tilde{C}_j\\
-{\rm e}^{{\rm i} \phi_i} \tilde{S}_i \tilde{C}_j
+{\rm e}^{{\rm i} \phi_j} \tilde{C}_i \tilde{S}_j&
\tilde{C}_i \tilde{C}_j+{\rm e}^{{\rm i}(\phi_i
-\phi_j)} \tilde{S}_i \tilde{S}_j\end{array} \right) \nonumber\\
\tilde{C}_i&=&\cos\frac{\theta_i}{2}\,,\,\,\,\,\tilde{S}_i=\sin \frac{\theta_i}{2}\,,
\label{eq:hop}
\end{eqnarray}
and $\theta_i, \phi_i$ are the polar co-ordinates of the spin
$\vec{S}_i$.

In the bulk of the ferromagnetic state, $t^{\alpha \beta}_{ij}$
reduces to a 
unit matrix, but inside the domain walls, the values of both diagonal and
off-diagonal elements are changed. Thus, the bandstructure (and hence 
the carrier density) within the wall differs from that in the bulk, and we come
to the conclusion that {\it Bloch walls are charged}. We will now
evaluate the surface charge $\sigma_B$ of a Bloch wall in a double
exchange ferromagnet. 

Let us suppose that the Bloch wall runs along the [11] direction of
the lattice diagonal, and choose the $y$ axis to be perpendicular to the wall.
We also choose the co-ordinates in spin-space in such a way that
$\phi_i \equiv 0$, and note that $\theta_i$ does not depend on $x$. In
other words, the spin configuration is composed of 
ferromagnetically ordered chains running in the $x$ direction, with
the interchain and intrachain distances given by $1/\sqrt{2}$ and
$\sqrt{2}$ respectively. It is then convenient to Fourier-transform
the fermion operators in the $x$ direction only, according to
\begin{equation}
d_\alpha(x,y)=\left(\frac{2}{N}\right)^{1/4} \sum_{k_x} {\rm e}^{{\rm i} k_x x
/\sqrt{2}}d_\alpha(k_x,y)\,,\,\,\,\,\,|k_x| < \pi\,. 
\label{eq:fourx}
\end{equation}
Then the first two terms in Eq. (\ref{eq:Ham1}) can be re-written in the
form 
\begin{eqnarray}
&&\!\!\!\!\tilde{\cal H}=\!-\!\!\!\! \sum_{y,k_x,\alpha , \beta}\!\!\! \cos{\frac{k_x}{2}}
\left[t^{\alpha \beta} (y,y+\frac{1}{\sqrt{2}})
d^\dagger_{\alpha}(k_x,y) d_{\beta}(k_x,y+\frac{1}{\sqrt{2}})+
\right. \!\!\!\!\!\nonumber \\
  &&\!\!\!\!+ {\rm
h.c.} \left]
+ \frac{J_H}{2}\right. \sum_{y,k_x} \left[ d^\dagger_{\downarrow}(k_x,y)
d_{\downarrow}(k_x,y) -
d^\dagger_{\uparrow}(k_x,y)
d_{\uparrow}(k_x,y)\right], \nonumber \\ && 
\!\!\!\!\!\!\!\!\!\!\!\!\!\!\!\label{eq:Ham2}
\end{eqnarray}
which we will also use in Sect. \ref{sec:abrupt} below.

In the ferromagnetic state, the subsequent Fourier transformation in
the $y$ direction according to

\begin{equation}
d_{\alpha}(k_x,y)=\left(\frac{1}{2N}\right)^{1/4} \sum_{k_y} {\rm
e}^{{\rm i} k_y y 
\sqrt{2}}d_{\vec{k}\alpha}\,,\,\,\,\,\,|k_y| < \pi\,. 
\label{eq:foury}
\end{equation}
yields the spectrum,
\begin{equation}
\epsilon^{\uparrow,\downarrow}_{\vec{k}} = \mp \frac{J_H}{2}
+\epsilon_{\vec{k}}\,,\,\,\,\,\, \epsilon_{\vec{k}}=-2\cos
\frac{k_x}{2} \cos k_y\,.  
\label{eq:fmspec}
\end{equation}

The variation of  spin direction within a Bloch wall corresponds to the
long-wavelength limit, $l_B \gg 1$, of continuum micromagnetic
theory. Then one can define $\theta(y)$ as a continuous function, and
the angle formed by the spins $\vec{S}(x,y)$ and $\vec{S}(x',y+
 1/\sqrt{2})$ on the neighbouring chains is given by $(\partial
\theta/\partial y)/\sqrt{2}$. For the case of a constant value of
$\partial \theta/\partial y \ll 1$, the spin-up fermion 
spectrum,
$\tilde{\epsilon}^{\uparrow}_{\vec{k}}=-(J_H/2)+\tilde{\epsilon}_{\vec{k}}$, 
is obtained from Eqs. (\ref{eq:hop}--\ref{eq:Ham2}) (upon  Fourier
transformation, Eq. (\ref{eq:foury})). 
When $\partial \theta/\partial y\neq 0$, the quantity $\tilde{\epsilon}_{\vec{k}}$
is only approximately factorisable,
\begin{eqnarray}
\tilde{\epsilon}_{\vec{k}}&=&\cos \frac{k_x}{2}\, (\epsilon_y(k_y)+ \delta
\epsilon_y(k_x,k_y))\,,\,\,\,\epsilon_y=-2 \cos k_y\,, \nonumber \\
\delta \epsilon_y &=& (\partial \theta/\partial y)^2 (\cos
k_y-\frac{4}{J_H}\cos \frac{k_x}{2} \sin^2 k_y )/8\,. 
\label{eq:deltaepsilon}
\end{eqnarray}
The value of carrier density at a fixed value of chemical potential is
then given by\cite{blochcharge} 
\begin{eqnarray}
&&n=x+\delta x= 
\\
&&=\int_{-\mu/2}^1 \frac{2 d \epsilon_x}{\pi \sqrt{1-\epsilon_x^2}}
\int_{-2}^{\mu/\epsilon_x} \left\{\rho_y(\epsilon_y) +
\delta \rho_y(\epsilon_x, \epsilon_y) \right\} d \epsilon_y.
\nonumber
\end{eqnarray}
Here, $\rho_y=1/(\pi\sqrt{4-\epsilon_y^2})$ is the value of density of
states at fixed $k_x$ in the ferromagnetic state, and $\delta \rho_y$ is the
correction arising at $\partial \theta/\partial y \neq 0$. Then the change in the
carrier density due to a non-zero value of $(\partial \theta/\partial
y) \ll 1$ can be evaluated (to leading order in $\delta \epsilon_y$) as
\begin{eqnarray}
\delta x &=& \int_{-\mu/2}^1 \frac{2 d \epsilon_x}{\pi
\sqrt{1-\epsilon_x^2}}
\delta n_y (\frac{\mu}{\epsilon_x})\,,\nonumber\\
\delta n_y
(\frac{\mu}{\epsilon_x})&\equiv&\int_{-2}^{\mu/\epsilon_x}\delta
\rho_y(\epsilon_x, \epsilon_y) d \epsilon_y \approx -\delta \epsilon_y
\rho_y(\frac{\mu}{\epsilon_x})\,.
\end{eqnarray}
Using Eq. (\ref{eq:deltaepsilon}), we obtain after some algebra
\begin{equation}
\delta x = \frac{\cal C}{2} \left( \frac{\partial \theta} {\partial
y}\right)^2\!,\,\,\,{\cal C}=\frac{1}{2 \pi^2} \left
[ \left(\frac{\mu}{4}-\frac{\mu^2}{2J_H} \right) Y_1(x) + 
\frac{2}{J_H} Y_2(x) \right].
\label{eq:deltax}
\end{equation}

Finally, given the known profile  of $\theta(y)$ in a Bloch
wall\cite{Volume8}, $\cos \theta(y)= {\rm \tanh}(y/l_B)$,
we find the following expression for the charge of a Bloch wall per
unit length:
\begin{equation}
\sigma_B= -e {\cal C}/l_B
\label{eq:blochcharge}
\end{equation}
where $e$ is the absolute value of electron charge. In evaluating
$\sigma_B$ as $-e \int \delta x dy$, we  used the
adiabatic approximation, which is valid in the long-wavelength limit of 
$l_B \gg 1$. As expected, a similar calculation for a Bloch wall
running parallel to a lattice direction yields the same result 
(\ref{eq:blochcharge}): Bloch walls have a well-defined continuum
limit, and both their energy\cite{energy} and charge are independent of the
orientation on a square lattice.

We note that at $J \geq 0$, the Bloch wall can be stable only as long
as the chemical 
potential at the centre of the wall (where the band-narrowing effect
is most pronounced) lies above the bottom of the carrier band. In
other words, the value of $x+\delta x$ with $\delta x$ given by
Eq. (\ref{eq:deltax}) should remain positive at $y=0$ (otherwise, there
would be no carriers and hence no carrier-mediated ferromagnetic interaction
near the centre of the wall)\cite{ferro}. Since the wall
is smooth, $l_B \gg 1$, this condition is important only at the
low-doping limit of $x \ll 1$, when it reads\cite{lowhole}
\begin{equation} 
16 \pi x DS > K\,.
\label{eq:blochstable}
\end{equation}
This is clearly violated at sufficiently low $x$. 
We will see that in this case the
domain wall is in fact abrupt [Sect. \ref{sec:abrupt},
Eq.(\ref{eq:abruptpreferred})]. 

According to Eq. (\ref{eq:blochcharge}), the charge of the Bloch wall,
which is inversely proportional to its width, decreases with
decreasing anisotropy strength: $\sigma_B \propto \sqrt{K}$. 
At small values of electron density $x\ll 1, J_H$, we find
$\sigma_B= e/(8 \pi l_B)$.
The behaviour of $\sigma_B$ at the intermediate doping levels can be
inferred from Fig. \ref{fig:bloch} (b), where the quantity ${\cal C}(x)$ (see
Eqs.(\ref{eq:deltax}--\ref{eq:blochcharge})) is plotted for different
values of $J_H$. We suggest that the
experimental determination of $\sigma_B$ may help to distinguish
Bloch walls from  abrupt or stripe domain walls (see Sections
\ref{sec:abrupt}--\ref{sec:stripe} below), which typically carry
larger charge. On the theory side, the effect of Bloch wall charge on
the carrier transport across the wall should be considered.

Throughout this section, we assumed\cite{omit} that the
Debye--H\"{u}ckel screening
radius is large 
in comparison to $l_B$. This appears to be  plausible,
especially in view of relatively large values of dielectric
constants, characteristic for the highly-polarisable oxides.
We will briefly  discuss the magnitude of Coulomb correction to
the Bloch wall energy, $S_B$, in Appendix \ref{app:coulomb} [Eq. (\ref{eq:blochcoulomb})].
In the opposite  case of strong screening, the charge of a Bloch wall
will vanish.

\section{ABRUPT WALL}
\label{sec:abrupt}

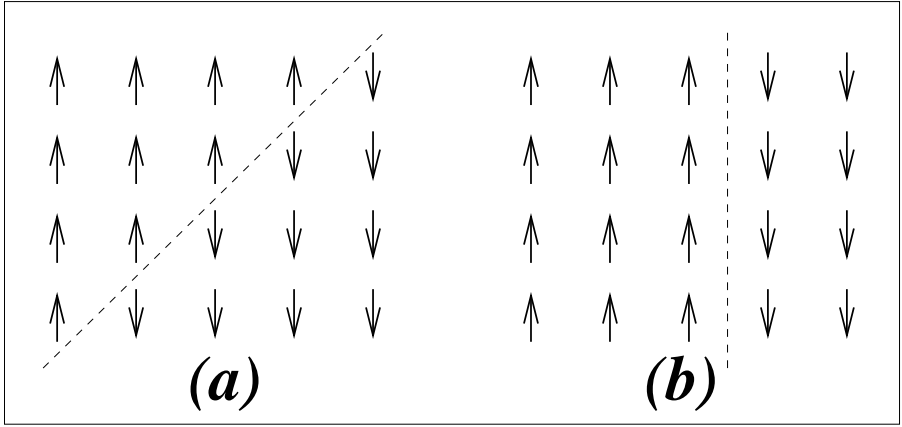
\begin{figure}
\caption{Diagonal {\it (a)} and vertical {\it (b)} abrupt domain walls
(dashed lines).}
\label{fig:abruptspins}
\end{figure}

The appreciable contribution of magnetic domain walls to resistivity,
as observed in certain ferromagnetic strained CMR
films\cite{Mathur99,Li,Wu},    
suggests the possibility of non-Bloch walls arising in these systems.
Indeed, in order to scatter the carriers effectively domain wall must
have a non-smooth structure, characterized by abrupt changes in spin 
direction. An abrupt (Ising-type) domain wall, shown in
Fig. \ref{fig:abruptspins}, represents an extreme example of such a structure.

Unlike the Bloch wall,  abrupt wall represents a lattice problem 
(as opposed to a long-wavelength one). Therefore the properties of an 
abrupt wall depend on its orientation with respect to the lattice, and one
has to distinguish between, {\it e.g.}, diagonal
(Fig. \ref{fig:abruptspins} a) and vertical
(Fig. \ref{fig:abruptspins} b) walls. We note that a similar
feature would also arise for domain walls in an Ising
ferromagnet -- indeed, the number of cut ferromagnetic links per
unit wall length is different for vertical and diagonal walls.
In a classical double exchange ferromagnet, the standard double exchange
mechanism forbids carrier hopping across
the abrupt
domain wall\cite{hundbound}. Owing to the anisotropy of the carrier
spectrum (as manifested in a non-spherical shape of the Fermi
surface), the carrier contribution to the 
abrupt wall energy is again orientation-dependent.  

In order to show that abrupt domain walls can actually arise in double
exchange ferromagnets, we will first turn to the low-doping limit, $x
\ll 1$, assuming also that $J_H=\infty$ and $J=0$. Since the Fermi momentum
is small, $p_F^2=4 \pi x \ll 1$,  carrier dispersion can be
approximated by the free-particle dispersion law,
$\epsilon^\uparrow_{\vec{k}} \approx {\rm const}+ (k^2/2)$.  
The energy of an abrupt wall is therefore equal to that of a partition
inserted into an ideal spin-polarised Fermi gas, which can be easily
estimated.

Let the ideal Fermi gas be contained in a rectangular box of the size
$L_x \times L_y$. According to the uncertainty principle (or alternatively to
the usual rules of momentum quantisation), the difference between the
allowed values $p_y^{(i)}$ of the y-component of momentum can be estimated as
$\delta p_y \sim 1/L_y$. Suppose now that a flat partition perpendicular
to the y axis has been introduced, dividing  the box in half. This
shifts each allowed momentum value: $p_y^{(i)}\rightarrow p_y^{(i)}+
\delta p_y^{(i)}$ with $|\delta p_y^{(i)}| \sim \delta p_y$. The signs
of $\delta p_y^{(i)}$ are chosen in such a way that the {\it energy}
shift of each individual electron level is positive: $\delta \epsilon
(p_x,p_y) \sim |p_y| \delta p_y$. The net energy change associated
with the partition is thus given by $L_x L_y \int n_{\vec{p}}
|p_y|d^2p/L_y$, or $ \sim x^{3/2}$ per unit length of partition\cite{3d}.
   
Thus, we find that the energy of abrupt domain wall in a double
exchange ferromagnet is given by $S_A \sim x^{3/2}$. The numerical
coefficient can be obtained by an exact treatment [see below and
Appendix \ref{app:iml},
Eqs. (\ref{eq:smallxdiag}--\ref{eq:smallxvert})], yielding $S_A \approx 4
\sqrt{\pi}x^{3/2}/3$. 
Comparing this with the Bloch wall energy, 
$S_B \approx \sqrt{Kx}$ (see
Eq.(\ref{eq:bloch})), we find that the abrupt wall energy is lower,
$S_A<S_B$, as
long as
\begin{equation}
x^2<9K/16 \pi\,.
\label{eq:abruptpreferred}
\end{equation}
We note that according to Eq. (\ref{eq:blochstable}), Bloch walls
become altogether unstable at $x^2<K/4\pi$.

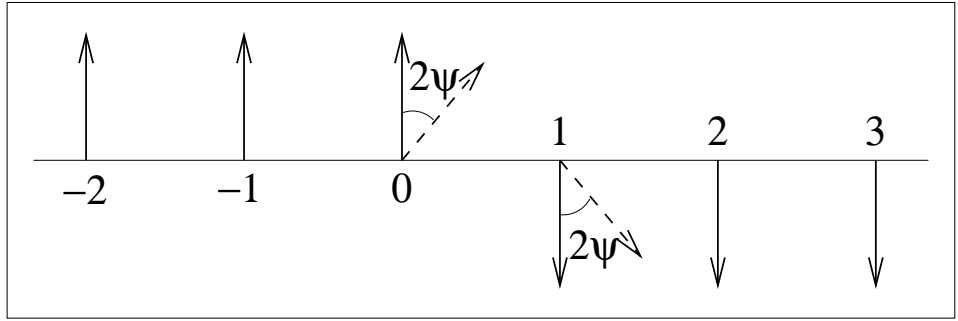
\begin{figure}
\caption{Schematic representation of a one-dimensional problem which
arises in diagonal domain wall calculations,
Eqs. (\ref{eq:Ham3}--\ref{eq:perturb}). The intersite
distance is equal to $1/\sqrt{2}$, and the numbers are the same as the
subscripts 
of the fermion operators in Eq. (\ref{eq:perturb}). Dashed arrows
correspond to the perturbed case, $\psi \neq 0$.}
\label{fig:abrupt1d}
\end{figure}

It appears to be very difficult to rigorously address the question
whether in the region specified by inequality
(\ref{eq:abruptpreferred}) the abrupt wall actually represents the
optimal spin configuration. We are, however, able to verify [see
Appendix \ref{app:iml},
Eqs. (\ref{eq:smallxdiag}--\ref{eq:smallxvert})] that as long as
$x^2<K/\pi$, 
the abrupt domain wall is stable 
with respect to small ``smearing'' perturbations (shown schematically
in Fig. \ref{fig:abrupt1d}) involving spins adjacent to the domain
wall on both sides. This provides a strong, albeit variational,
argument for the overall stability of abrupt walls.

We now turn to exact calculation of energies and charges of abrupt
walls for all values of $x$, $J_H$, and $J$, beginning with the
evaluation of the electronic contribution to the energy of an abrupt
diagonal wall. 

Following the Fourier transform, Eq. (\ref{eq:fourx}), the electronic
terms in the Hamiltonian of the uniform ferromagnetic phase take the
form (cf. Eq. (\ref{eq:Ham2}))
\begin{eqnarray}
&&\tilde{\cal H}=\sum_{k_x} {\cal H}_{k_x} \,,\\
&&{\cal H}_{k_x}=-\frac{Q}{2} \sum_y \left\{d^\dagger_\uparrow
(k_x,y+\frac{1}{\sqrt{2}}) d_\uparrow(k_x, y) +\right. \nonumber \\
&&+\left. d^\dagger_\downarrow
(k_x,y+\frac{1}{\sqrt{2}}) d_\downarrow(k_x, y)  +{\rm h.c.} \right\}+
\nonumber \\ 
&& 
+ \frac{J_H}{2} \sum_{y} \left\{ d^\dagger_{\downarrow}(k_x,y)
d_{\downarrow}(k_x,y) -
d^\dagger_{\uparrow}(k_x,y)
d_{\uparrow}(k_x,y)\right\}\,, 
\label{eq:Ham3}
\end{eqnarray}
where $Q=2 \cos (k_x/2)$. The abrupt diagonal domain
wall parallel to the $x$ axis results in a perturbation of the
Hamiltonian (\ref{eq:Ham3}), ${\cal H}_{k_x} \rightarrow {\cal
H}_{k_x} + V_{k_x}$, with
\begin{eqnarray}
\frac{2}{Q}V_{k_x}&& = \left\{d^\dagger_{-1 \uparrow} d_{0\uparrow} + 
d^\dagger_{-1 \downarrow} d_{0\downarrow} +d^\dagger_{1 \uparrow}
d_{2\uparrow} +  
d^\dagger_{1 \downarrow} d_{2\downarrow} \right\}\times \nonumber \\
&&\times (1- \cos \psi) +
\left\{d^\dagger_{0 \uparrow} d_{1\uparrow} + 
d^\dagger_{0 \downarrow} d_{1\downarrow} \right\} ( 1- \sin 2 \psi)
+ \nonumber \\
&&+\left\{d^\dagger_{0 \uparrow} d_{1\downarrow} - 
d^\dagger_{0 \downarrow} d_{1\uparrow} \right\} \cos 2 \psi +
\left\{d^\dagger_{-1 \uparrow} d_{0\downarrow} - 
d^\dagger_{-1 \downarrow} d_{0\uparrow}+\right. \nonumber \\
&&\left. +d^\dagger_{1 \uparrow}
d_{2\downarrow} -  
d^\dagger_{1 \downarrow} d_{2\uparrow} \right\} \sin \psi
+ {\rm h. c.}\,.
\label{eq:perturb}
\end{eqnarray}
Here we denoted $d_{\alpha}(k_x,i/\sqrt{2})$ by $d_{i \alpha}$ and
allowed for a smearing perturbation, $\psi \ll 1$, as shown in
Fig. \ref{fig:abrupt1d}. It is convenient to re-write the operator
$V_{k_x}$ in a diagonal form,
\begin{equation}
V_{k_x}=\sum_{i=1}^{8}A_i a^\dagger_i a_i\,,\,\,\,\,a^\dagger_i a_j+
a_j a^\dagger_i= \delta_{ij}\,. 
\label{eq:diag}
\end{equation}
Expressions for both the eigenvalues $A_i$ and the operators $a_i$ are
given in Appendix \ref{app:iml}.

In the absence of a domain wall, the electronic contribution to 
thermodynamic potential of a double exchange ferromagnet at a
temperature $T$ can be
evaluated as
\begin{eqnarray}
\Omega & =& \int \frac{L_x dk_x}{2 \pi \sqrt{2}} \int \frac{L_y \sqrt{2} d
k_y}{2 \pi} \varphi[\epsilon_{\vec{k}}]= \nonumber \\ 
&=&\int \frac{L_x dk_x}{2 \pi
\sqrt{2}} \int d \epsilon \nu_{tot} (\epsilon, Q) \varphi(\epsilon)
d\epsilon\,, \nonumber\\
\epsilon_{\vec{k}} &=& -Q \cos k_y\,,\,\,\,\,\, \varphi(\epsilon)= -T
{\rm ln} \left[1+ {\rm exp} \left(\frac{\mu - 
\epsilon}{T}\right)\right] \,.
\label{eq:Omega0}
\end{eqnarray}
Here, $L_x$ and $L_y$ are the dimensions of the sample,
$\nu_{tot}(\epsilon,Q)= L_y \sqrt{2}/(\pi \sqrt{Q^2-\epsilon^2})$ is the
total density of states at a fixed value of $Q$ [{\it i.e.,} with 
$k_x= \pm 2{\rm arccos}
(Q/2)$], and the factors $\sqrt{2}$ originate in momenta rescaling
implied in Eqs. (\ref{eq:fourx}) and (\ref{eq:foury}).
  
When the domain wall perpendicular to the $y$ axis is introduced, the
associated perturbation $V_{k_x}$,
Eqs. (\ref{eq:perturb}--\ref{eq:diag}), gives rise to a correction\cite{1d} in
the density of states, $\nu_{tot}(\epsilon,Q) \rightarrow
\nu_{tot}(\epsilon,Q) + \delta \nu (\epsilon, Q)$. Introducing the
Lifshits--Krein spectral shift function \cite{iml} $\xi(\epsilon, Q)$
according to $\delta \nu = - \partial \xi /\partial \epsilon$, we find
for the electronic contribution to the domain wall energy,
\begin{eqnarray}
\frac{\delta \Omega}{L_x} &=& \int \frac{dk_x}{2 \pi \sqrt{2}} \int d
\epsilon \delta \nu (\epsilon, Q) \varphi(\epsilon)=\nonumber \\
& =& \int\frac{dk_x}{2 \pi \sqrt{2}} \int d
\epsilon \xi(\epsilon, Q) f(\epsilon)\,.
\label{eq:trace1}
\end{eqnarray}
Here, the zero-temperature value for the Fermi distribution function,
$f(\epsilon) = \theta(\mu-\epsilon)$, can be substituted.    

For a given value of $k_x$, the operator $V_{k_x}$ represents a local
perturbation of a one-dimensional Hamiltonian ${\cal H}_{k_x}$. Thus, the 
dependence of $\xi$ on $Q$ is only parametric\cite{1d}, and the value
of $\xi$ can be found from the standard formula\cite{iml} (see also
Ref. \cite{prb98}):
\begin{equation} 
\xi\big(\epsilon,Q(k_x)\big) = - \frac{1}{\pi} {\rm Arg\, Det}
\left[ \hat{1}- \hat{G}(\epsilon-\frac{1}{2}J_H -{\rm i} 0, Q) V_{k_x} \right].
\label{eq:xigeneral}
\end{equation}
where $\hat{G}(\zeta,Q)=(\zeta \cdot \hat{1}-{\cal H}_{k_x})^{-1}$ is the
resolvent operator at a given value of $k_x$, and $\hat{1}$ is the
identity operator. In the basis
containing the states $a_i^\dagger |0\rangle$ ( where $|0\rangle$ is
the vacuum state), the determinant on the r.\ h.\ s. of
Eq. (\ref{eq:xigeneral}) is that of an 8 $\times$ 8 matrix,         
$\delta_{ij}-M_{ij} A_j$, with
\begin{equation}
M_{ij}= \sum_{\alpha=\uparrow,\downarrow} \int \frac{dk_y \sqrt{2}}{2
\pi Q} \frac{\langle 0 |a_i |k_y^\alpha \rangle \langle k_y^\alpha |
a_j^\dagger | 0 \rangle}{E_\alpha + \cos k_y - {\rm i} 0}\,.
\label{eq:mij}
\end{equation}
Here
\begin{equation}
E_\uparrow = \epsilon/Q\,,\,\,\,\,\,\, E_\downarrow= (\epsilon -J_H)/Q\,,
\label{eq:Eupdown}
\end{equation} 
and $|k_y^\alpha \rangle$ are properly normalised Bloch wave states,
\begin{eqnarray}
|k_y^\alpha\rangle &=& \frac{1}{2^{1/4}}\sum_y {\rm e}^{{\rm -i}k_y y \sqrt{2}}
 d^\dagger_\alpha (k_x,y) |0 \rangle \,,\label{eq:wavediag}\\
 \langle k_y^\alpha | k_y^{\prime\, \beta}
 \rangle &=& 2 \pi \delta(k_y-k_y^\prime) \delta_{\alpha\beta}.
\nonumber
\end{eqnarray}

After a straightforward, if somewhat laborious, calculation we obtain
\begin{eqnarray}
&&\xi(\epsilon,Q)=\xi^{(0)}(\epsilon,Q) + \delta \xi,\,\,\,{\rm tg} \pi
\xi^{(0)} = \frac{E_\uparrow E_\downarrow
-1}{\sqrt{E^2_\downarrow-1}\sqrt{1-E^2_\uparrow}}\,, 
\nonumber\\
&&\delta \xi=\frac{4 J_H^2}{\pi Q^2}\sqrt{1-E_\uparrow^2}\,
\frac{E_\uparrow-E_\downarrow-\sqrt{E_\downarrow^2-1}}{E_\downarrow^2-E_\uparrow^2-2E_\uparrow
E_\downarrow} \,\psi^2\,.
\label{eq:xifinal}
\end{eqnarray}
The final expression for the energy of an abrupt diagonal domain wall
per unit length is then given by the trace formula,
Eq. (\ref{eq:trace1}), with additional contributions from direct
superexchange and single-ion anisotropy:
\begin{eqnarray}
&&S_d\equiv S_d^{(0)}+Z_d \psi^2 = -2\sqrt{2}J+2 \sqrt{2}(2J+K)
\psi^2+ \nonumber \\
&&+\frac{\sqrt{2}}{\pi}\int_{-1}^{1}dE_\uparrow
\int_{0}^{2}\frac{QdQ}{\sqrt{4-Q^2}}   
\,\xi(QE_\uparrow,Q) \theta(\mu-QE_\uparrow)\,.
\label{eq:diagabrupt}
\end{eqnarray}
The energy of a vertical abrupt domain wall is calculated very
similarly (see Appendix \ref{app:iml}), yielding the result
\begin{eqnarray}
&&S_v\equiv S_v^{(0)}+Z_v \psi^2=-2J+4(J+K)\psi^2+ \nonumber \\
&&+\frac{1}{\pi}\int_{-1}^{1}\frac{d\epsilon_1}{\sqrt{1-\epsilon_1^2}}
\int_{-1}^{1}d\epsilon_2  \tilde{\xi}(\epsilon_2)
\theta(\mu-\epsilon_1-\epsilon_2)\,.
\label{eq:vertabrupt}
\end{eqnarray}
Here, $\tilde{\xi}(\epsilon_2)$ is equal to $\xi(\epsilon_2,Q)$ as given by
Eqs. (\ref{eq:xifinal}) with 
$E_\uparrow= \epsilon_2$, $E_\downarrow = \epsilon_2 -J_H$, and Q=1.

The spectral shift function, Eq. (\ref{eq:xifinal}), also contains information
about the abrupt domain wall charges. Indeed, spectral shift
function $\xi(\epsilon)$ generally measures the number of energy
levels that cross
the given energy value $\epsilon$ as a result of a  perturbation. 
Thus, the change in electron
density at a fixed value of $k_x$ is given by $-\xi(\mu,Q)$, yielding
the charge of an unperturbed ($\psi=0$) abrupt diagonal wall:
\begin{equation}
\sigma_d=\frac{\sqrt{2}}{\pi} e \int_{|\mu|}^{2} \xi^{(0)} \left
( \frac{\mu}{Q}\,, Q \right) \frac{dQ}{\sqrt{4-Q^2}}\,.
\label{eq:sigmadiag}
\end{equation}
For a vertical wall, we likewise obtain
\begin{equation}
\sigma_v=\frac{e}{\pi} \int_{|\mu|-1}^{1}\tilde{\xi}^{(0)}(\mu -
\epsilon_1 {\rm sgn 
\mu}) \frac{d \epsilon_1}{\sqrt{1-\epsilon_1^2}}\,.
\label{eq:sigmavert}
\end{equation}
Here the function $\tilde{\xi}(\epsilon)$ is defined in the same way as in
Eq. (\ref{eq:vertabrupt}) above.

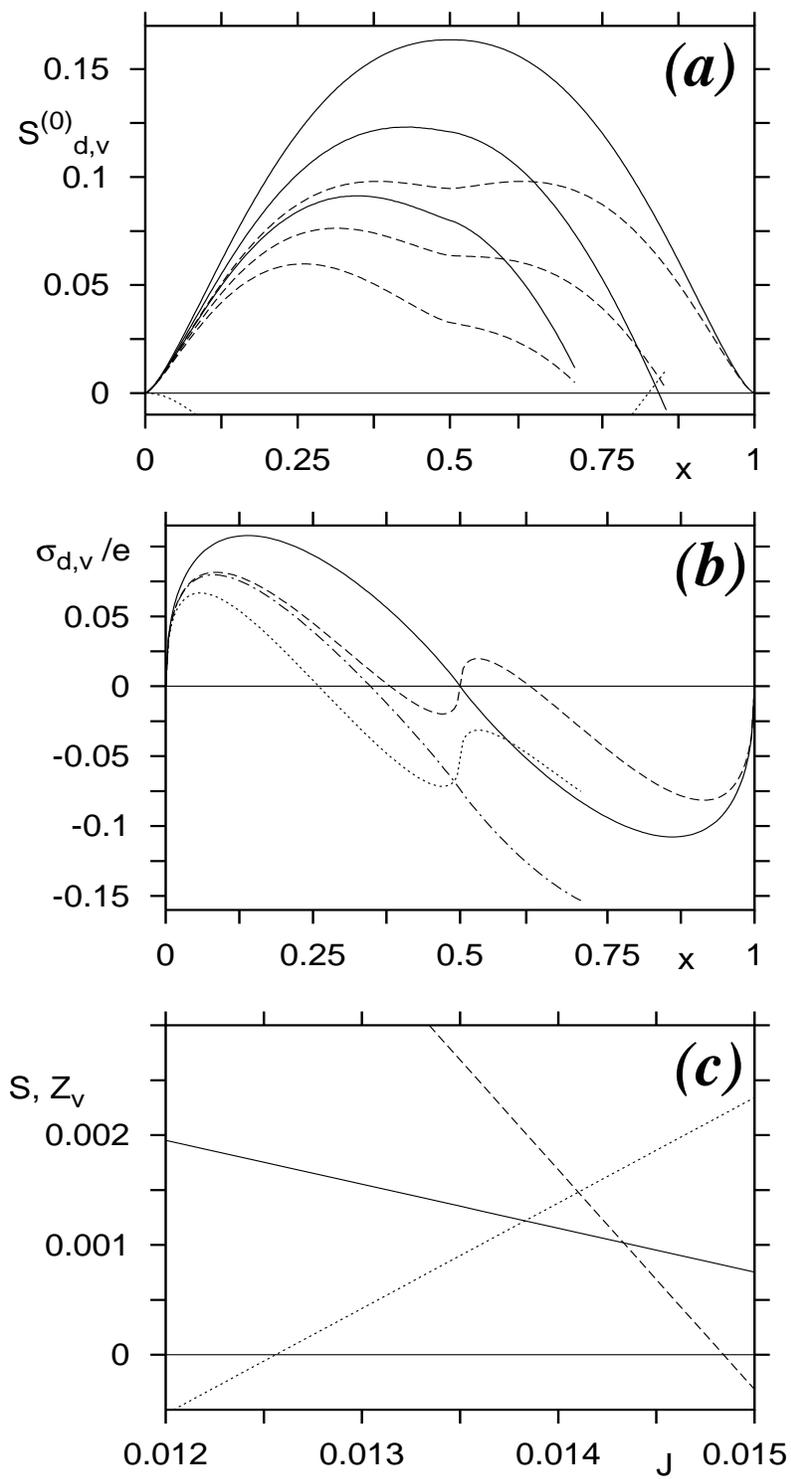
\begin{figure}
\caption{{\it (a)}. Abrupt wall energies vs. $x$ at $J=0$. Solid
(dashed) lines, top to bottom: diagonal wall energy, $S^{(0)}_d$
(vertical wall energy, $S^{(0)}_v$) for $J_H \rightarrow \infty$,
$J_H=8$, and $J_H$=4. For finite values of $J_H$, the lines end at the
values of $x$ corresponding to the sign change of spin stiffness,
$D$. Immediately below these values, the ferromagnetic state is
unstable with respect to phase separation (see
Fig. \ref{fig:bloch}). For $K(x)=D(x)S/25$, ({\it i.e.,} 
$l_B=5$) the quantities $Z_{d,v}$ are negative 
everywhere outside the low-doping regions $x \ll 1$ and $1-x \ll 1$, except for
the case of $J_H=8$, when $Z_d$ becomes positive for
$x>0.83$ (dotted line).  {\it (b)}. Abrupt wall charges in units of
electron charge, $|e|$. Solid and dashed (dashed-dotted and dotted) 
lines represent $\sigma_d$ and $\sigma_v$ for $J_H \rightarrow \infty$
($J_H=4$). {\it (c)}. Bloch wall energy $S_B$ (solid line), abrupt
vertical wall energy $S_v^{(0)}$ (dashed line), and the quantity $Z_v$
(dotted line) vs. superexchange $J$. Anisotropy constant varies
according to $K(J)=DS/25$. Conduction electron density and Hund's rule
coupling strength are given by $x=0.55$ and $J_H=4$, respectively, and
the system is unstable with respect to phase separation.}
\label{fig:abruptnumerics}
\end{figure}

We have conducted a thorough numerical investigation of
Eqs. (\ref{eq:diagabrupt}--\ref{eq:vertabrupt}). Doping dependence of
the abrupt wall energies for different values of $J_H$ is illustrated
in Fig. \ref{fig:abruptnumerics} {\it (a)}. Comparing these with
Fig. \ref{fig:bloch} {\it (a)}, we conclude that at the intermediate
doping levels, abrupt wall energies are typically several times larger
than spin
stiffness, $DS$.
Therefore in the physically relevant case of small anisotropies, $K
\ll DS$, Bloch walls will typically have a significantly lower energy
[see Eq. (\ref{eq:bloch})]. We note that including antiferromagnetic
superexchange, $J>0$, would lead to a decrease in $S^{(0)}_d$ relative to
$S^{(0)}_v$ [as follows form
Eqs. (\ref{eq:diagabrupt}--\ref{eq:vertabrupt})]. In particular, this
can yield\cite{unstabletyp} $S^{(0)}_d<S^{(0)}_v$ at small values of $x$.

The charges, $\sigma_d$ and $\sigma_v$, of the abrupt domain walls are
plotted in Fig. \ref{fig:abruptnumerics} {\it (b)}. We see that at the
intermediate doping values, the
electric charge per unit length is of the order of 
$0.1 e$, in a marked difference from weakly-charged Bloch walls
[cf. Fig. \ref{fig:bloch} and Eq. (\ref{eq:blochcharge})].

With increasing antiferromagnetic interactions in the system (that is,
either with increasing $J$ or with decreasing $J_H$) spin stiffness,
as well as the abrupt wall energies, will eventually change
sign. Near this point, there might be a region where $DS$ is still
positive, while either $S_v^{(0)}$ or $S_d^{(0)}$ is smaller than the
Bloch wall 
energy, $S_B$. This is due to the fact that $S_B$ and $S_{v,d}^{(0)}$ depend
on $J$ and $J_H$ in different ways. Such a situation is illustrated in
Fig. \ref{fig:abruptnumerics} {\it (c)}, showing the domain wall
energies in a double exchange ferromagnet with $x=0.55$ and $J_H=4$ as
functions of $J$. The solid line corresponds to the Bloch wall energy
$S_B$, whereas the the vertical abrupt wall
energy, $S_v^{(0)}$, is represented by a
dashed line. The value of easy-axis anisotropy constant, $K$, is
varied with $J$ in such a way that Bloch wall width, $l_B$ [see
Eq. (\ref{eq:bloch})], is always equal to 5. We see that
$S_B>S_v^{(0)}>0$ for $0.0143<J<0.0148$. Since the quantity $Z_v$ 
[see Eq. (\ref{eq:vertabrupt})],
represented by the dotted line is positive throughout the $S_v^{(0)}<S_B$
region, one is tempted to conclude that the abrupt wall is indeed
stable in this region. However, it is easy to verify that for $J>0.0107$,the
homogeneous ferromagnetic ground state becomes unstable with respect
to phase separation into ferro- and antiferromagnetic regions. 
It appears that this represents the general situation, {\it i.e.},
that at the intermediate doping range the inequality $S_B>S_v^{(0)}$ (or
$S_B > S_d^{(0)}$) cannot be satisfied within the thermodynamically
stable region. 
In Sect. \ref{sec:stripe} below, we will argue that the phenomenon of phase 
separation can affect the magnetic domain wall structure in a profound
way. Here we merely note that even if phase separation is suppressed
due to some mechanism ({\it e. g.}, enforcing electric neutrality on
the microscopic level), the parameter region where either $S_v$ or
$S_d$ is smaller than $S_B$ (but the stiffness $D$ is still positive)
would still be very narrow, requiring one to fine-tune the values of
$J$, $J_H$, $K$, 
and $x$ to within a fraction of a per cent\cite{intermediate}. It is
therefore very
unlikely that such a situation can be realized experimentally in a
reproducible way.  

Expressions (\ref{eq:diagabrupt}--\ref{eq:sigmavert}) can be further
simplified in the limiting cases of small carrier density, $x \ll 1$,
or large Hund's rule coupling, $J_H
\rightarrow \infty$ (see Appendix
\ref{app:iml}). Expressions
(\ref{eq:diagabruptinfj}--\ref{eq:sigmavertinfj}), valid in the $J_H
\rightarrow \infty$ limit, can be used to estimate the values of $S_d$
and $S_v$ at sufficiently large $J_H$ throughout the entire range of 
dopant concentrations.

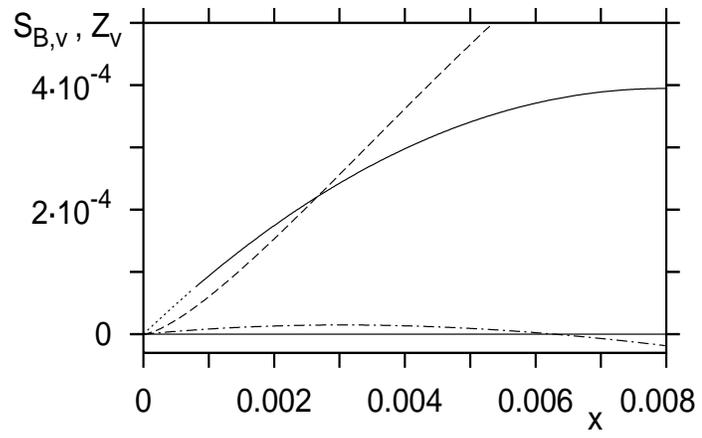
\begin{figure}
\caption{Bloch wall energy $S_B$ (solid line), abrupt
vertical wall energy $S_v^{(0)}$ (dashed line), and the quantity $Z_v$
(dashed-dotted line) vs. electron density, $x$, in the low-density
limit without superexchange (J=0). 
Hund's rule coupling is fixed at $J_H=0.1$, while the
anisotropy varies according to $K(x)=DS/25$, leading to a constant Bloch
wall width, $l_B=5$. The Bloch wall, however, becomes unstable
at lower $x$ (dotted line). }
\label{fig:smallx}
\end{figure}

As discussed in the beginning of this section [see
Eq. (\ref{eq:abruptpreferred})] , the domain walls become
abrupt at the low-doping limit of $x\ll 1$. In this case, 
the abrupt wall energies and charges are given by
Eqs. (\ref{eq:smallxdiag}--\ref{eq:smallxsigmavert}). The doping
dependence of domain wall energies in this region is illustrated
in Fig. \ref{fig:smallx}. The value of Hund's rule coupling is taken
to be $J_H=0.1$, and $K(x)=D(x)S/25$ again ensuring that $l_B=5$.
We see that the the energy of an abrupt vertical wall (dashed line) is
lower than that of a Bloch wall (solid line), $S_v^{(0)}<S_B$, for all
$x<0.0027$,  
and the stability of abrupt domain
wall is further evidenced by the fact that the the quantity $Z_v$
(dashed-dotted line) is positive for $x<0.0063$. For this choice of
parameters\cite{J0},
the value of $S_d^{(0)}$ is just above that of $S_v^{(0)}$, and we
find
 $S_d^{(0)}<S_B$ at $x<0.0026$, and $Z_d>0$ for all $x<0.0074$. 
Bloch wall becomes unstable [see Eq. (\ref{eq:blochstable})] at
$x<0.0008$ (dotted line). We note that
allowing for a larger value of $K$ would have broadened the region
where abrupt walls have lower energy; however, Eqs. (\ref{eq:bloch})
are valid only in the $l_B \gg 1$ case.

The data shown in   
Fig. \ref{fig:smallx} are for a system with no direct superexchange,
$J=0$; including a small $J>0$ would give rise to a phase-separation
instability at small $x$ \cite{jap02}, which may or may not cover the
entire region of $S_{v,d}^{(0)}<S_B$. While no study of domain structure
in the electron-doped manganates has been reported so
far, it appears that superexchange in these systems is sufficiently
strong to destabilise the homogeneous ferromagnetic
state at $x\ll 1$\cite{Maignan}. The abrupt wall picture as discussed
here is then inapplicable (see Sect. \ref{sec:stripe}
below). We note, however, that this might not 
be the case 
for other lightly doped magnetic semiconductors or semimetals. Ferromagnetic
semiconductors such as ${\rm Eu}$-doped ${\rm EuS}$ and ${\rm Eu O}$
have relatively high values of Curie temperature $T_C$
(Ref. \cite{nagaevreview}), presumably originating from a
strong ferromagnetic superexchange, $J<0$. In this case, even in a
lightly-doped sample
ferromagnetism is due mostly to superexchange (rather than to double
exchange) and one expects that the domain walls will be of Bloch type,
like in conventional Heisenberg ferromagnets. However, other magnetic
semiconductors such as ${\rm Eu Se}$  
become ferromagnetic only upon small electron doping
\cite{shapira}. In this case of
small positive $J$, domain walls may in fact be abrupt. This also may
be the case in a ferromagnetic
semimetal ${\rm EuB}_6$ (Ref. \cite{Ott}).
It would therefore be most
interesting to study experimentally the domain wall structure (in
particular, the effect of domain walls on the transport properties) in
the ferromagnetic films of these compounds.  

Throughout our calculation, we neglected the effects of chemical disorder
which can lead to localisation of electron states. We note that the overall
profile of carrier wave functions does not directly affect the properties
of an abrupt wall. The assumption essential for our approach is that the
electron wave function can be {\it locally} approximated by an energy 
eigenfunction of the clean case\cite{plane} with the same energy. 
This is valid provided
that the localisation length is much larger than the inverse Fermi momentum;
the latter condition is expected to be satisfied in manganates within the
metallic regime, as well as in the doped magnetic semiconductors and semimetals
discussed above.  

\section{PHASE SEPARATION AND STRIPE WALLS}
\label{sec:stripe}

Phase separation is a phenomenon which commonly occurs in the CMR
manganese oxides\cite{Nagaevbook,dagottoreview}. Although  direct evidence is
lacking, it appears likely that the films studied in
Refs. \cite{Mathur99,Li,Wu}  are in fact phase-separated. It is
therefore important to consider the effect of phase separation on
magnetic domain wall structure in double exchange ferromagnets.

Let us first suppose that the values of parameters of the system (that is,
carrier density $x$, superexchange $J$, Hund's rule coupling strength
$J_H$) lie within the stability region of the uniform ferromagnetic
phase. The thermodynamic potential  is then given by
$\Omega_{FM}=E+2J-\mu x$, with the value of $\mu=\mu_{FM}(x)$
determined by the uniform conduction electron density $x$. The electron
charge density, $-ex$, is compensated by the combined charge of
magnetic and non-magnetic ions,
resulting in  electric neutrality of the system on
the microscopic level\cite{singleband}. As the values of parameters are varied ({\it
e.g.,} either the value of $J$ is increased or that of $J_H$ is
decreased), the system  eventually becomes unstable
with respect to phase separation into ferromagnetic phase and another
phase which we will call antiferromagnetic\cite{zoo}. In the absence
of Coulomb interaction, this occurs  when the
thermodynamic potentials of the two phases become equal to each other:
$\Omega_{FM}(\mu_{FM}(x))=\Omega_{AFM}(\mu_{FM}(x))$. At this point,
it becomes energetically advantageous to create islands of the
antiferromagnetic phase within the bulk of 
ferromagnet. Since there is a finite energy cost $W$ 
associated with a unit length of the boundary between the two phases,
such an island should contain a large number of sites in order to reduce the
boundary energy per antiferromagnetic site; as long as this
is the case, the area occupied by the antiferromagnetic phase can be 
arbitrarily small relative to the total size of the system, so that 
the carrier density $x$ within the ferromagnetic area
and hence the value of chemical potential $\mu_{FM}(x)$ remain unchanged. 

Structure of the boundaries between different phases has been studied
by the present writer in Ref. \cite{jap02}. It was found that at least
in some cases these 
boundaries are abrupt; it appears plausible that this property is rather
generic. We note that the energy and charge of an abrupt interphase
boundary can be evaluated using the approach applied in
Sect. \ref{sec:abrupt} above to the study of abrupt domain wall. A
boundary between ferro- and antiferromagnetic areas can be perfectly
abrupt only if it runs parallel to certain lattice
directions\cite{jap02}. It is therefore likely that within a large
region of parameter values, the emerging islands of antiferromagnetic
phase will have a square (or diamond) shape. Apart from one case
discussed towards the end of this section, the latter feature is
unimportant for the rather qualitative discussion below.
We will therefore assume that the islands are circular, which would
correspond to the boundary  energy $W$ independent on direction.

While the chemical potential $\mu=\mu_{FM}(x)$ is constant across the
sample, the carrier density within the island, $x_{AFM}$, is different
from the nominal value $x$. We note that phase separation consists precisely in
a redistribution of the carriers with a simultaneous change in magnetic
ordering, and would not be possible had the requirement of constant
carrier density been enforced on the microscopic level. The island is
therefore electrically charged, and it is imperative to take into account the
effects of electrostatic Coulomb interaction and screening on phase
separation. 

In a thin film, the inverse Debye--H\"{u}ckel screening  radius is given
by\cite{Stern,Fetter} (see also Appendix \ref{app:coulomb}):
\begin{equation}
\kappa=\frac{2 \pi e^2
\nu_0}{\bar{\epsilon}}\,,\,\,\,\bar{\epsilon}=
\frac{1}{2}(\epsilon_{d1}+\epsilon_{d2})\,.  
\label{eq:2ddebye}
\end{equation}
Here $\nu_0$ is the value of carrier density of
states at the Fermi level and $\epsilon_{d1}$,$\epsilon_{d2}$ are
dielectric constants of the media on both 
sides of the conducting layer.  In the 3D case, which is 
discussed in more detail in Appendix \ref{app:coulomb},
$\kappa_{(3D)}^2=4 \pi e^2 \nu_0/\epsilon_d$, where $\epsilon_d$ is the
dielectric constant of the double exchange magnet itself.
If the size of the island was large in
comparison with Debye--H\"{u}ckel radius, $R \gg
\kappa^{-1}$, screening within the island would have restored
the carrier density to its nominal value $x$ (and charge density to
zero). In the case when there is no conduction band in the bulk
antiferromagnetic phase ({\it e.g.}, when $x_{AFM}(\mu)$ equals either
$0$ or $1$), the presence of electric potential $\varphi$ (which in
this case is strongly position-dependent) would shift the carrier band
within the island either
upwards or downwards. This in turn will ultimately give rise to a
Fermi surface,
screening, and restoration of the carrier density to its nominal value
on the length scale of $\kappa^{-1}$. 
However, as explained above, when the value of density is fixed no
phase separation is possible.
We therefore conclude that formation of an island can be
energetically favourable only as long as\cite{2Dscreen} $R
\stackrel{<}{\sim}\kappa^{-1}$. We will assume for simplicity that
$1\ll R
\ll \kappa^{-1}$, that is, that the carrier density within the island
is uniform and equal to the bulk value of $x_{AFM}(\mu)$. This obviously
includes an assumption that Debye--H\"{u}ckel radius is large on the
atomic length scale, $\kappa \ll 1$. The latter is not unphysical, 
in view of relatively large dielectric constants $\epsilon_d$ reported for the
manganates\cite{alexandrov} 
and of suppression of carrier density of states at the Fermi level
$\nu_0$ found in the $x$-ray absorption and angle-resolved
photoemission measurements\cite{dos}. For a thin film, the situation also
depends on the choice of the substrate, as discussed in more detail in the
end of this section.

With these assumptions, the change in thermodynamic potential
$\Omega$ associated with a creation of a single circular antiferromagnetic
island in a 2D system can be evaluated as\cite{omitcharge}
\begin{eqnarray}
X_i&=&\pi R^2 (\Omega_{AFM}-\Omega_{FM})+ 2 \pi R W + \frac{1}{2} \int d^2r
\rho(\vec{r}) \varphi(\vec{r}) + \nonumber \\
&&+\int_{(FM)} d^2r \int_\mu^{\mu^\prime} d \epsilon
(\epsilon -\mu) \nu_0\,.
\label{eq:loneisland1}
\end{eqnarray}
Here, the first two terms represent the bulk and boundary
contributions, the third term is the electrostatic Coulomb energy, and
the last term is the kinetic energy cost of re-distributing electrons
in the ferromagnet, caused by the shift of electrochemical potential
$\mu^\prime(\vec{r})=\mu+e \varphi(\vec{r})$ (that is, the shift of
band energies due to the presence of electric field within the
screening cloud). Charge density $\rho(r)$ equals 
$\rho_{AFM}=-e(x_{AFM}-x)$ within the island, and $-e \delta
x(\vec{r})$ outside, where $\delta x$ is the change of electron density
in the screening cloud.  The last term in Eq. (\ref{eq:loneisland1})
can be re-written as 
\[\frac{1}{2} \int_{(FM)}d^2r \frac{(\delta x)^2}{\nu_0}  =
\frac{1}{2} \int_{(FM)}e \varphi \delta x d^2r = - \frac{1}{2} \int_{(FM)}
\rho \varphi d^2 r \,.\]
This allows us to render Eq. (\ref{eq:loneisland1}) in the form
\begin{equation}
X_i=\pi R^2 (\Omega_{AFM}-\Omega_{FM})+ 2 \pi R W + \frac{1}{2}
\int_{(AFM)} d^2r 
\rho\varphi
\label{eq:loneisland2}
\end{equation}
where the integration in the last term is carried out over the area of
the island. Evaluating the potential
$\varphi$ to leading order in  $1/(\kappa
R)\ll 1$ [see Appendix \ref{app:coulomb},
Eqs. (\ref{eq:loneisland2.1}--\ref{eq:loneisland2.2})], we obtain 
\begin{equation}
X_i=\pi R^2 (\Omega_{AFM}-\Omega_{FM})+ 2 \pi R W + \frac{8 \pi
\rho_{AFM}^2 R^3}{3 \bar{\epsilon}}\,.
\label{eq:loneisland3}
\end{equation}
Creation of an island becomes energetically favourable once the
minimum value of this expression drops below zero. This
yields the 
following threshold condition for the phase separation to occur:
\begin{equation}
\Omega_{FM}-\Omega_{AFM}> \Delta_0=8 |\rho_{AFM}| \sqrt{\frac{W}{3
\bar{\epsilon}}}
\label{eq:threshold}
\end{equation}
[at $\Omega_{FM}=\Omega_{AFM}+\Delta_0$, the discriminant of the cubic
equation $X_i(R)=0$ vanishes; the minimum value, $X_i(R_0)=0$, is then
reached at $R_0=(3\bar{\epsilon}W)^{1/2}/(2|\rho_{AFM}|)$].

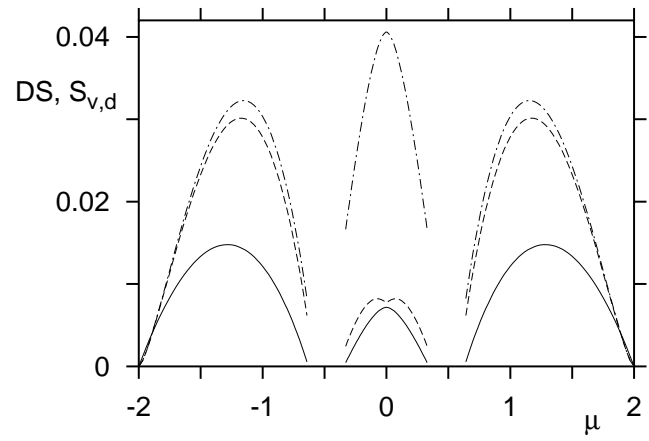
\begin{figure}
\caption{Chemical potential dependence of spin stiffness $DS$ (solid
line) and the 
diagonal and vertical abrupt domain wall energies (dashed-dotted and
dashed lines) for a $J_H \rightarrow \infty$ system on the brink of
phase separation. The value of $J$ is adjusted in such a way that
$\Omega_{FM}=\Omega_{AFM}$ for any value of carrier density $x$. The
nature of corresponding antiferromagnetic phases is discussed in the text.}
\label{fig:balance}
\end{figure}

Let us now consider a domain wall in a phase-separated film. We note
that in this case the antiferromagnetic and ferromagnetic tendencies
in the system are approximately balanced against each other; this
greatly reduces both the spin stiffness [which in turn determines the
Bloch wall energy via Eq. (\ref{eq:bloch})] and the energy of abrupt
domain walls, $S^{(0)}_{v,d}$. This point is illustrated by
Fig. \ref{fig:balance}, representing the chemical potential dependence of spin
stiffness (solid line) and abrupt wall energies (dashed and
dashed-dotted lines) for a
$J_H\rightarrow \infty$ system with the value of $J=J(\mu)$ adjusted in
such a way\cite{smallcorr} that $\Omega_{FM}=\Omega_{AFM}$. The appropriate 
antiferromagnetic
phase near the endpoints $\mu=\pm 2$ is characterized by the usual
N\'{e}el $\{ \pi, \pi\}$  (G-antiferromagnetic) spin ordering, whereas
in the vicinity of quarter-filling, $\mu=0$, the A-antiferromagnetic phase
with the ordering vector $\{\pi,0\}$ proves more advantageous. The plethora
of possible phases arising in the intermediate case (see
Ref. \cite{jap02}) are not considered here, and no value is plotted for
$DS$ and $S^{(0)}_{v,d}$ unless the phase separation into the
ferromagnetic and either G- or A-antiferromagnetic phases is possible.
Comparing Fig. \ref{fig:balance} with the $J=0$ case, plotted in
Figs. \ref{fig:bloch} {\it (a)} and 
\ref{fig:abruptnumerics} {\it (a)}, we find a drastic
reduction of  both spin stiffness and domain wall energies at the
intermediate doping values. In
addition, the energies of abrupt domain walls are now of the same
order of magnitude as spin stiffness, in a marked difference with the
single-phase case considered earlier. 

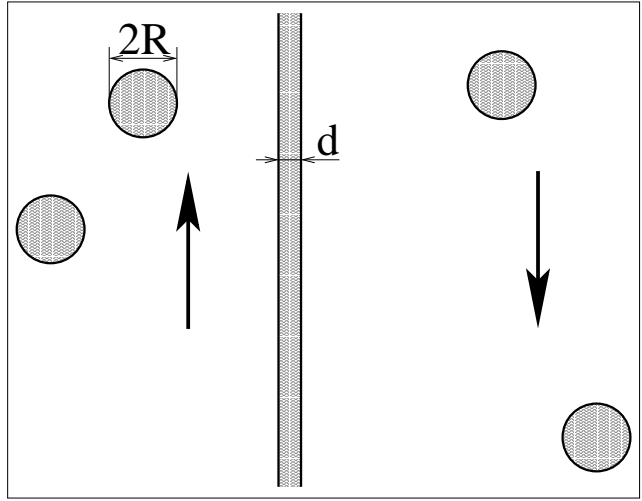
\begin{figure}
\caption{Schematic representation of a stripe domain wall in a
phase-separated double exchange magnet. The two ferromagnetic domains
with antiparallel directions of magnetisation (arrows) are separated
by a stripe of antiferromagnetic phase (shaded). In addition,
unconnected islands of antiferromagnetic phase are formed within each domain.}
\label{fig:stripe}
\end{figure}

We will first discuss the effect of Coulomb forces in the case when the
value of $\Omega_{FM}-\Omega_{AFM}$ is just above the threshold,
Eq. (\ref{eq:threshold}), so that the islands of antiferromagnetic
phase arising within each ferromagnetic domain are well separated from
each other, and Eqs. (\ref{eq:loneisland1}--\ref{eq:loneisland3}) are
valid. As discussed in Sect. \ref{sec:abrupt}, the
abrupt domain wall shuts the carrier hopping in the perpendicular
direction, acting as a partition in the gas of conduction electrons. 
In the absence of Coulomb forces, the energy cost of creating a stripe
of antiferromagnetic phase adjacent to the wall is therefore equal to
$-(\Omega_{FM}-\Omega_{AFM})d$ (where $d$ is the stripe width) per unit
length of the stripe, and does not include any additional boundary
contribution. This statement (which is equivalent to saying that the
abrupt wall energy is equal to $2W$ per unit length) is exact in the
$J_H \rightarrow \infty$ limit (see Appendix \ref{app:iml} and
Fig. \ref{fig:infjscheme}). It is also clear that it provides a
reasonable estimate for the case of large but finite $J_H$; the
details of situation at finite $J_H$ will be addressed elsewhere.
Thus, when $\Omega_{FM}-\Omega_{AFM}>0$, the energy of an abrupt
domain wall can be further lowered by inserting alongside it a stripe of
antiferromagnetic phase (see Fig. \ref{fig:stripe}). The width of the
stripe is determined by a trade-off between the bulk and Coulomb
energies, {\it i.e.}, by minimising the energy of a {\it stripe domain wall}
per unit length,
\begin{equation}
S_s(d)=2W-(\Omega_{FM}-\Omega_{AFM})d-\frac{d^2}{\bar{\epsilon}}\rho_{AFM}^2
{\rm ln}\kappa d
\label{eq:lonestripe}
\end{equation}
[see Appendix \ref{app:coulomb},
Eqs.(\ref{eq:string}--\ref{eq:stringtextbook})]. Since the
antiferromagnetic stripe separates two ferromagnetic 
domains with antiparallel directions of magnetisation, the spins at
the two edges of the stripe must point in the opposite
directions. For the stripe of $A$-antiferromagnet
($G$-antiferromagnet) parallel to a lattice direction (lattice
diagonal), this means that the number $d$ (the number $\sqrt{2}d$)
must be odd\cite{abrupt}; similar conditions 
should hold for other phases. Since we assumed that the value of $d$
is sufficiently large, $d\gg 1$, these requirements do not affect our
estimates. Assuming that
$\Omega_{FM}-\Omega_{AFM}= \Delta \Omega_0$, 
we find 
\begin{equation}
d_s=\frac{8 \sqrt{\bar{\epsilon} W/3}}{\rho_{AFM} {\rm ln} 
\frac{\rho_{AFM}^2}{\kappa^2 \bar{\epsilon} W}}\,,\,\,\,
S_s(d_0)\approx 2W-\frac{16}{3} W \frac{1}{{\rm ln} \kappa d_s}.
\label{eq:stripewall1}
\end{equation}

Eqs. (\ref{eq:lonestripe}--\ref{eq:stripewall1}) are valid to leading
order in $\kappa d \ll 1$; even though ${\rm ln} \kappa d_s$ is thus large,
the relatively large coefficient of $16/3$ in the second term of
Eq. (\ref{eq:stripewall1}) allows for a significant reduction of 
domain wall energy due to the presence of a stripe of
antiferromagnetic phase. It is not impossible that this reduction
can make the quantity $S_s$ lower than the Bloch wall energy $S_B$,
provided that the easy-axis anisotropy constant $K$ is sufficiently large.
The domain walls would then have a stripe structure, and would
strongly interfere with the transport properties of the system.
However, the exact values of quantities $\kappa$ and $W$ in
Eq. (\ref{eq:stripewall1})  are not known, and it is not
clear whether this situation can be realized experimentally.

More importantly, Eq. (\ref{eq:stripewall1}) [and its 3D analogue,
Eq. (\ref{eq:stripewall13D})] refer to the case when phase separation
is just beginning, with the islands of antiferromagnetic phase
separated by large areas of a ferromagnet. Indeed, our derivation
relied on an assumption that the screening clouds formed around
different antiferromagnetic islands do not overlap, that is, that the
inter-island distance is much larger than the screening radius. The
size of each island, on the other hand, is much smaller than
$\kappa^{-1}$, so only a small part of the net sample area is occupied 
by the antiferromagnetic phase, making phase separation difficult to
detect. The available experimental data on phase separation in the CMR
compounds\cite{dagottoreview}, on the other hand, correspond to the
case when a substantial part of the sample reverts to a
non-ferromagnetic phase.  Within the context of phase separation
mechanism considered here this is only possible when neither the
size of antiferromagnetic islands (or stripes) nor the inter-island
distance is larger than Debye--H\"{u}ckel radius. Below we will
consider the case when screening is negligible (that is, when the
inter-island distance is much smaller than $\kappa^{-1}$). Since
$\kappa$ is expected to be small (see above), this is not unrealistic;
moreover, the results are expected to provide a reasonable estimate
for the case of intermediate screening strength as well.    

The ferro- and antiferromagnetic phases are then characterized by
uniform values of electron densities $x_{FM}$ and $x_{AFM}$ and charge
densities, $\rho_{FM}=-e(x_{FM}-x)$ and $\rho_{AFM}=-e(x_{AFM}-x)$.
The numbers of sites occupied by ferro- and antiferromagnetic phases,
\begin{equation}
N_{FM}=\frac{N}{1+\delta}\,,\,\,\,N_{AFM}=\frac{N\delta}{1+\delta}\,,
\,\,\,\,\delta \equiv - \frac{\rho_{FM}}{\rho_{AFM}} 
\end{equation} 
(where $N$ is the total number of sites in the system) are self-adjusted in
such a way that the values of bulk thermodynamic potentials of the two
phases, $\Omega_{FM}$ and $\Omega_{AFM}$, are close to each other. 
Therefore our observation that both spin stiffness $DS$ and abrupt
wall energies are significantly reduced and are of the same order of
magnitude (see above and Fig. \ref{fig:balance}) remains applicable.

It is expected that the value of parameter $\delta$ can be determined
experimentally.
 
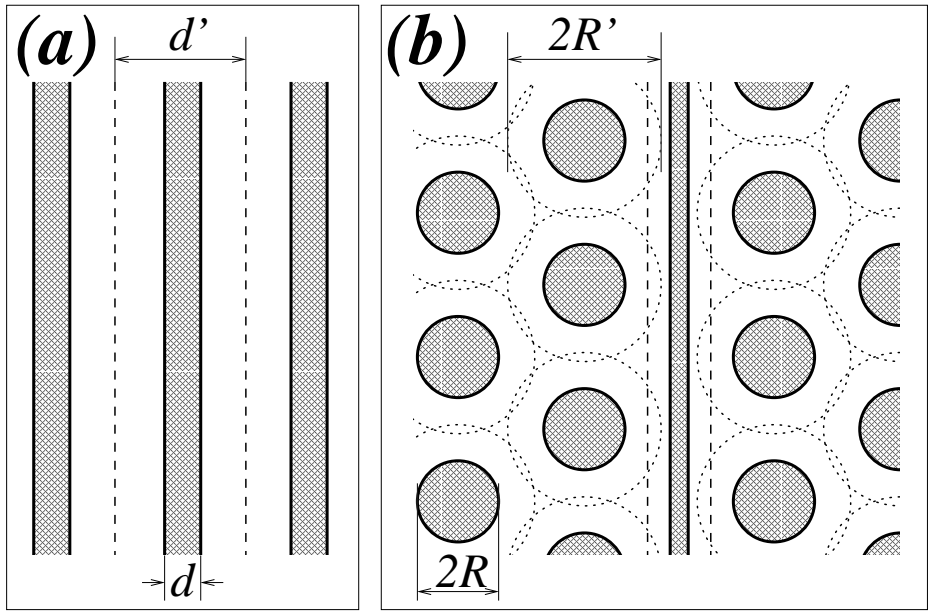
\begin{figure}
\caption{Stripe phase  {\it (a)} and stripe domain wall within the
droplet phase {\it (b)}. 
The system is phase-separated into 
ferromagnetic (unshaded) and antiferromagnetic (shaded) regions with 
$\delta \approx 0.4$.
The Wigner cell boundaries of stripe and droplet phases are shown with
dashed and dotted lines respectively. The two connected ferromagnetic
domains extending to the left and to the 
right of the stripe wall in {\it (b)} are magnetised in the opposite
directions (not shown). The width of antiferromagnetic stripes in {\it
(a)} and {\it (b)} is given 
by Eqs. (\ref{eq:stripephasewidth}) and (\ref{eq:stripe2width}), 
respectively.}
\label{fig:stripe2}
\end{figure}

We are interested in the situation when within each ferromagnetic
domain the poorly-conducting antiferromagnetic phase forms
disconnected droplets (so that metallic conductance through the
connected ferromagnetic area is still possible), and we will again
assume that these droplets are circular in shape. The number of
droplets in the sample is then $N_{AFM}/(\pi R^2)$ (where
$R$ is the radius of a droplet), and  thermodynamic potential of
the phase-separated system is given by
\begin{equation}
\Omega_1(R) = \frac{\Omega_{FM}+\Omega_{AFM} \delta}{1+\delta}+
\frac{1}{\pi R^2} \frac{\delta}{1+\delta} \left( 2 \pi R W + E_1\right)
\label{eq:omega1.1}
\end{equation}
per site, where $E_1$ is the Coulomb energy of a single droplet. This
term cannot be evaluated rigorously; in order to estimate it, we
calculate the energy of Coulomb interaction within the so-called
Wigner cell, composed of the droplet and a surrounding ring
$R<r<R^\prime$ (where $r$ is the distance from the centre of the
droplet) of the ferromagnetic phase. The value of
$R^\prime=R[(1+\delta)/\delta]^{1/2}$ is chosen in such a way that the
combined charge of the droplet and the ring vanishes. It should be
emphasised that unless $\delta$ is small, $\delta \ll 1$, this
procedure, which has been used to treat a similar 
problem earlier\cite{Nagaev74}, is not exact\cite{exact3d}: even though
the electrostatic potential of a Wigner cell falls off rapidly with
distance, $\varphi(r) \propto r^{-3}$, it does not vanish outside the
cell. In addition, different 
Wigner cells overlap with each other. Thus, by using this approach we 
essentially replace the Coulomb force with some model interaction, 
which however captures the essential features of the original problem
as long as the value of $\delta$ is not too large
(see below). We find $E_1=8 \pi (R^\prime)^2 R \rho_{AFM}^2 A_1(\delta)
\delta/(3 \bar{\epsilon})$ [see Appendix \ref{app:coulomb}, Eq. (\ref{eq:a1delta})], where for
small values of $\delta \ll 1$ the function $A_1(\delta)$ is equal to 1.
The thermodynamic potential of the droplet phase, Eq. (\ref{eq:omega1.1}),
has to be minimised with respect to the droplet radius $R$, yielding
\begin{equation}
\Omega_1 = \frac{\Omega_{FM}+\Omega_{AFM} \delta}{1+\delta}+
\frac{8|\rho_{FM}|}{\sqrt{1+\delta}} \sqrt{\frac{A_1(\delta) W}{3
\bar{\epsilon}}}\,.
\label{eq:omega1.2}
\end{equation}
Another possible geometry of phase separation is represented by the
stripe phase [shown in Fig. \ref{fig:stripe2} {\it (a)}], formed by 
the parallel antiferromagnetic stripes of width $d$ embedded into the
ferromagnetic background.
The thermodynamic potential of the stripe phase is given by
\begin{equation}
\Omega_2(d) = \frac{\Omega_{FM}+\Omega_{AFM} \delta}{1+\delta}+
\frac{1}{d} \frac{\delta}{1+\delta} \left( 2  W + E_2\right)\,,
\label{eq:omega2.1}
\end{equation}
Within the Wigner-cell
approximation, Coulomb energy per unit
length of a single stripe, $E_2$, is calculated by subdividing the
sample into the ``Wigner stripes'' of width
$d^\prime=d(1+\delta)/\delta$ [see Fig. \ref{fig:stripe2} {\it (a)}].
We find $E_2=-(d d^\prime \rho_{AFM}^2 A_2(\delta)
\delta/\bar{\epsilon}) {\rm ln \delta}$ with $A_2(\delta \rightarrow
0)=1$ [see Appendix \ref{app:coulomb},
Eq. (\ref{eq:a2delta})]. Minimising the value of 
$\Omega_2$
with
respect to $d$, we obtain 
\begin{eqnarray}
d_0&=&\frac{1}{|\rho_{AFM}|} \sqrt{\frac{2 W
\bar{\epsilon}} {(1+\delta) A_2(\delta)| {\rm ln} \delta|}}\,,
\label{eq:stripephasewidth} \\
\Omega_2-\Omega_1 &=& \frac{|\rho_{FM}|}{\sqrt{1+\delta}}
\sqrt{\frac{W}{\bar{\epsilon}}} \left( 2 \sqrt{2 A_2(\delta) |{\rm ln}
\delta|}- 8 \sqrt{\frac{A_1(\delta)}{3}}  \right)\,.
\nonumber
\end{eqnarray}

The latter quantity is positive for all values of $\delta$ between $0$ and
$1$, indicating that within this model approach, the droplet phase is
always preferred (see below). The
formation of a stripe domain wall in the droplet phase involves
re-arranging spins 
within a Wigner stripe of width $d^\prime_s=d_s(1+ \delta)/\delta$
into the stripe phase [see 
Fig. \ref{fig:stripe2} {\it (b)}], that is, forming a single stripe of
antiferromagnetic phase [of width $d_s(\delta)$] flanked by two stripes
of ferromagnet. The net area occupied by antiferromagnetic phase,
$N_{AFM}$, is conserved, as is the overall electric neutrality.
Minimising the stripe wall energy per unit length,
$S_s=(\Omega_2(d_s)-\Omega_1)d^\prime_s$, with respect to $d_s$, we find 
\begin{eqnarray}
d_s&=&\frac{4}{ A_2(\delta) |\rho_{AFM}{\rm ln} \delta|}\sqrt{ 
\frac{W \bar{\epsilon} A_1{\delta}}{3(1+\delta)}} \,, 
\label{eq:stripe2width} \\
S_s&=&2WB(\delta)\,, \,\,\,\,B(\delta)=1-\frac{8}{3}\frac{A_1(\delta)}
{A_2(\delta) |{\rm ln} \delta|}\,.
\label{eq:stripe2}
\end{eqnarray} 
The ratio, $B(\delta)$, of the stripe wall energy $S_s$ to the abrupt
wall energy, $2W$, is plotted in Fig. \ref{fig:stripeplot}
(solid line). We see that the inclusion of an antiferromagnetic stripe 
can lower the energy of an abrupt wall by a factor of $4$. Since
the spin stiffness $DS$ is of the same order as the abrupt wall energy 
(see Fig. \ref{fig:balance}), the stripe wall energy can be lower than
the Bloch wall energy $S_B$ already at a moderate value of anisotropy,
$K \sim DS/64$ [cf. Eq. (\ref{eq:bloch})].

Within the Wigner-cell approach for circular droplets
the other droplet phase, with the ferromagnetic droplets in the
antiferromagnetic background, becomes preferred at $\delta >1$
(cf. Ref. \cite{Nagaev74}).
While this transition might give rise to new possible domain wall
structures near  $\delta =1$, this is not expected to be physically
relevant due to the intrinsic limitations of the method. As the value
of $\delta$ increases towards unity, the Wigner cell estimate for
Coulomb energy becomes progressively less reliable due to 
decreasing separation between the droplets. It is perhaps even more
important that the effects of the droplet shape can no longer be ignored.

As mentioned above, it is likely that the optimal shape of
antiferromagnetic droplets is square; this would be in line with
earlier results for double exchange model\cite{jap02}, as well as with
the numerical results for phase 
separation in other similar systems\cite{Berkovits}. 
In order to calculate the energy of the square-droplet phase at small
$\delta$, one can still use the Wigner-cell approach. Due to the
increase in the droplet boundary energy, the combined Coulomb and
boundary contribution to the thermodynamic potential
of the droplet phase [the last term in Eq. (\ref{eq:omega1.2})]
increases by some 
6 \%. This in turn leads to a noticeable decrease in the quantity
$B(\delta)$ (dotted line in Fig. \ref{fig:stripeplot}).

As the value of $\delta$ increases, the
Wigner-cell method becomes completely unsuitable for the analysis of
the square-droplet phase. Indeed, at $\delta=1$ (that is, at
$N_{FM}=N_{AFM}$) the square-droplet phase corresponds to a
checkerboard arrangement of equal ferro- and antiferromagnetic
squares, which has nothing in common with the Wigner cell picture
(cf. Fig. \ref{fig:stripe2}).
It is therefore clear that thermodynamic potential of the square-droplet
phase at sufficiently large $\delta$ is well above the value given by
Eq. (\ref{eq:omega1.2}). Accordingly, Eq. (\ref{eq:stripe2})
significantly over-estimates the value of $B(\delta)$ and hence the
stripe wall energy, $S_s$. 
While leaving this subject for future investigation, we note that it is
entirely possible that at a certain value of $\delta=\delta_c<1$ 
thermodynamic potential of the square-droplet phase exceeds that of
the stripe phase, Eq. (\ref{eq:omega2.1}). The quantities $B(\delta)$
and $S_s$
will vanish at this point\cite{point}, as exemplified schematically by the
dashed-dotted line in Fig. \ref{fig:stripeplot}. In this case,
for any finite value of the anisotropy constant $K>0$ and sufficiently
small $\delta_c-\delta>0$, magnetic domain walls within the conducting
phase  would have stripe (as opposed to Bloch-like)
structure.

We close with a brief comment on the applicability of our analysis
to the finite-thickness films. The results of
Sections \ref{sec:bloch} and \ref{sec:abrupt} for the spin stiffness
and abrupt wall energies are valid only as long as the carrier
velocity component perpendicular to the film is negligible. However,
our conclusion that on the brink of phase separation $DS$ and the abrupt
wall energies are generally of the same order of magnitude 
(as illustrated by Fig. \ref{fig:balance}) is likely to remain valid
in 3D as well. Our assumption that screening has  the two-dimensional
character is valid as long as
the film thickness remains small in comparison with the 
two-dimensional Debye--H\"{u}ckel radius, $\kappa^{-1}$. The film is
then thin from the viewpoint of electrostatics
[cf. Eq. (\ref{eq:potential2d})], that is, there is no electric field in the 
perpendicular direction within the film\cite{normal}. The latter holds provided 
that the film itself is homogeneous in this direction, {\it i.e.,}
that characteristic length scale
of a phase-separated sample (the droplet radius, $R \sim
(\bar{\epsilon} W)^{1/2} /|\rho_{AFM}|$), is larger than the film
thickness. 
Given the typical
experimental observations\cite{dagottoreview} that phase separation occurs
on the scale of at least 50-100 nm, this last condition is not particularly 
restrictive.

The Debye radius can be roughly estimated by assuming that $\nu_0$ is
of the order of inverse bandwidth ($4t \sim 5 {\rm  eV}$) divided by
the unit cell area ($\sim 0.15 {\rm nm}^2$). Taking in
Eq. (\ref{eq:2ddebye}) $\epsilon_{d2}=1$
(dielectric constant of the air), we then find $\kappa^{-1} \sim
(\epsilon_{d1}+1) \cdot 0.08 {\rm nm}$. The substrate used in the
measurements of Ref. \cite{Li}, lanthanum aluminate, has the
dielectric constant\cite{LAO} of $\epsilon_{d1} \approx 24$, resulting
in $\kappa^{-1} \sim 2 {\rm nm}$. It is therefore tempting to associate the
reported domain wall resistance\cite{Li} (large for the thinnest  ${\rm Pr_{2/3}Sr_{1/3}MnO_3}$ films of 4 nm,
vanishing for films thicker than 20 nm), which is observable below the
Curie temperature, $T_C \approx 130K$, with the stripe walls which
arise only as long as the the thickness of conducting layer (which is
presumably somewhat thinner that the film itself) is not large\cite{thick3D} in
comparison with $\kappa^{-1}$. We note that the film thickness
required for the lattice periods (and hence the anisotropy constant,
$K$, and Bloch wall energy) to relax to their bulk values is of the
order of 500 nm (cf. Ref. \cite{Wu}). Thus, our suggestion
provides an (hitherto lacking) interpretation for the disappearance of 
domain wall resistivity in the films thicker than only 20 nm.

The experiments of Ref. \cite{Mathur99},
on the other hand, were performed with (ferroelectric) strontium
titanate substrate, with\cite{STO} $\epsilon_{d1} \approx 1200$ at
$T=110 K$, which yields $\kappa^{-1} \sim 100 {\rm nm}$. The  ${\rm
La_{0.7}Ca_{0.3}MnO_3}$ (with the Curie temperature $T_C=250K$) film\cite{Mathur99} was 200 nm
thick, and the domain wall contribution  was observable below
$T=110K$. Given the strong dependence of $\epsilon_{d1}$ on temperature
($\epsilon_{d1} \approx 24,000$  at low $T$, $\epsilon_{d1} \approx
300$ at room temperature), it appears plausible that domain walls have
stripe structure at low temperatures, when the film thickness is not
large in comparison with $\kappa^{-1}$. Furthermore, it is not
unlikely that the above-mentioned transition at $T\approx 110K$ is due
to the violation of this condition at larger $T$, and associated change of the
domain wall structure\cite{structure}. We emphasize that this discussion is 
speculative at best, as we make no attempt to adequately describe the crossover
between two- and three-dimensional screening nor to take into account
the peculiar geometry of the sample used in Ref. \cite{Mathur99}.

It appears that stripe wall formation is in principle also
possible in the opposite limiting case of a bulk 3D material, although the 
Wigner-cell estimates given in Appendix \ref{app:coulomb} suggest that
somewhat higher values of $\delta$ are required. The values of Debye
radius, $\kappa_{(3D)}^{-1}$, and dielectric constant, $\epsilon_d$,
of doped manganates are, however, not known, and, crucially, very
small values of anisotropy make the Bloch wall energy very low.
It is therefore expected that in the 3D case the energy of Bloch wall is
generally lower than that of a stripe wall, in agreement with the fact
that no observable domain wall contribution to resistivity was
reported for the
manganate crystals.

\section{DISCUSSION}
\label{sec:conclu}

In this article we showed that there are at least three different
possible types of  structure of a ferromagnetic domain wall,
{\it all of which can be realized within the double exchange
model}. The energies and charges of Bloch, abrupt, and stripe domain
walls are also different, as are their anticipated contributions to
the resistance and magnetoresistance of the sample. The conventional, weakly 
charged Bloch walls (Sect. \ref{sec:bloch}), which generally arise in
single-phase samples, become unstable at low carrier densities, when the
abrupt walls (Sect. \ref{sec:abrupt}) are preferred. For a phase-separated
system, however, there is a region of parameter values when the domain
walls acquire stripe structure (Sect. \ref{sec:stripe}), characterized
by a stripe of antiferromagnetic phase separating the two domains.  

It is not yet
known whether all three types of wall can 
occur in the CMR manganate compounds. As follows from the discussion
in Sect. \ref{sec:abrupt},  abrupt walls are expected to arise
at low values of electron doping\cite{hole}, $x\ll 1$, provided that the 
homogeneous ferromagnetic phase remains
thermodynamically stable. We are not aware of any measurements of
the domain wall contribution to transport in this regime, and it is
not clear whether such a situation (which also requires the value of
direct superexchange $J$ to be extremely small) can
be realized in the manganates (however, see the end of
Sect. \ref{sec:abrupt} for a discussion of other compounds). 
As for the intermediate doping values, it
appears that domain walls can have either stripe or Bloch
structure. 

The effect of Bloch walls on the charge transport
properties of a double exchange ferromagnet has been discussed
theoretically\cite{Yamanaka}. The results are consistent with simpler
estimates \cite{Mathur99,Li,Wu} suggesting that for a realistic value
of $l_B$ and at an intermediate doping level, carrier
scattering off the Bloch wall cannot possibly account for a measured
domain wall contribution to the resistivity of the systems studied in
Refs.\cite{Mathur99,Li,Wu}. Measurable domain wall contributions to
the transport properties of the CMR manganates,  reported in other
studies known to us, are
attributable to the grain boundary effects in polycrystalline films 
\cite{Gupta,Miller,Mathur97}. In this case, the magnetic 
structure\cite{Soh2000,Miller} 
of a domain wall arising at a substrate grain boundary is largely
determined by underlying lattice defects\cite{Soh2002}. It is anticipated that
this also holds for the magnetic pattern appearing in a strained film 
at the boundary of a heavy-ion 
irradiated region\cite{Vlasko2002}. We note that the effects of lattice 
irregularities of any type are not included in the present theoretical 
treatment. 
  
Our results suggest
that magnetic domain walls in monocrystals or epitaxial films of CMR
manganates at the intermediate doping levels generally have Bloch-like
structure, with a notable exception of 
certain strained films similar to those used in 
Refs. \cite{Mathur99,Li,Wu}. Regarding the latter case, we expect that
domain walls may in fact be the stripe walls introduced in Sect
\ref{sec:stripe} above. This suggestion is corroborated by 
especially strong effect reported in Ref. \cite{Li}, which shows
that domain walls give a dominant contribution to the resistivity of
a thin ${\rm Pr_{2/3}Sr_{1/3}MnO_3}$ film at low temperatures. The
connexion between domain wall resistivity and dielectric properties
of the substrate, discussed in the end of Sect. \ref{sec:stripe},
appears to lend further support to the stripe wall scenario. 
The  stripe walls appear likely to arise in this case due to the strain-induced
increase of easy-axis anisotropy constant $K$ (which 
in turn increases the Bloch wall energy), and also to phase separation
which makes formation of the stripe walls possible. While it is not
clear whether phase separation does occur in the samples used in
Refs. \cite{Mathur99,Li,Wu}, this would be rather plausible given that
phase separation is commonly observed in both manganate crystals and
films\cite{dagottoreview}. We suggest that further measurements ({\it
e. g.}, scanning tunnelling spectroscopy) need
to be carried out to clarify whether these samples are in fact
phase-separated. On the other hand, domain wall properties
(including possible domain wall contribution to the resistivity) of
those CMR films which {\it are} known to 
phase-separate\cite{dagottoreview,Nosov}  
should also be investigated. Synthesis of electron-doped
manganate films, if technologically possible, may represent a promising new
direction\cite{Maignan}. We note that magnetic domain walls appear
only when a substantial fraction of the film is in the ferromagnetic
state, allowing for a low-field metallic conduction. 

In the present article, we did not quantitatively address the problem
of conduction across a domain wall of either type. The
available theoretical estimates of
domain wall conductance (Ref. \cite{Yamanaka} for Bloch walls,
Ref. \cite{Zhang} for abrupt wall) are incomplete in that the Coulomb
interaction between the carrier and the (charged) domain wall is not
taken into account. As for the stripe walls, the issue of
magnetotransport in this case has yet to be treated theoretically,
although it is clear that stripe wall contribution to resistivity is
much larger than that of either Bloch or abrupt walls.
In the presence of stripe domain walls, magnetoresistance will be
affected by the change of their structure under  a magnetic field,
which is likely to include a field-driven transition from stripe to
Bloch walls. It is therefore expected that the dependence of the domain wall
contribution to resistivity on the magnitude of applied in-plane field
can be different for the Bloch and stripe cases (smooth decrease for
Bloch walls, as opposed to possibly step-like features for stripe
walls, as seen in Ref. \cite{Mathur99}). 

Magnetotransport studies are not the only way to investigate the
properties of magnetic domain walls. Direct probes
of charge and spin structure of domain walls are possible in
principle (cf. Ref. \cite{Wiesendanger}), but have not yet been
performed for the manganates. However, Fresnel imaging
measurements on a thin ${\rm La_{0.7}Ca_{0.3}MnO_3}$ film were
reported recently\cite{Mathur01}. Domain walls were found to retain a
finite width of the order of 40 nm, in apparent agreement with
Eq. (\ref{eq:bloch}) for Bloch walls. We hope that domain wall widths 
in the strained films studied in Refs. \cite{Mathur99,Li,Wu}
will also be measured in the near future. It would be most interesting to
try to relate these to the band structure, magnetic, and electrostatic 
properties of the corresponding compounds and to check the agreement
with the estimates (\ref{eq:stripewall1}) and (\ref{eq:stripe2width})
for the stripe walls.  

\acknowledgements

The author takes  pleasure in thanking A. Berger, R. Berkovits, L. Brey, 
Y. F. Hu, K. Levin, P. B. Littlewood,
N. D. Mathur, M. R. Norman, Y.-A. Soh, V. K. Vlasko-Vlasov, U. Welp, 
M. E. Zhitomirsky, and especially J. T. Chalker 
for enlightening and motivating discussions. This
work was supported by EPSRC under grant GR/M04426.

\appendix
\section{Derivation and Analysis of Eqs. (\REF).}
\label{app:iml}

The key step in the calculation of the spectral shift function,
Eq. (\ref{eq:xifinal}), is the diagonalisation of perturbation
operator, $V_{k_x}$ [see
Eqs. (\ref{eq:perturb}) and (\ref{eq:diag})]. Its eigenvalues $A_i$ and
the corresponding fermionic operators $a_i$ are given by
\begin{eqnarray}
A_1=A_2=-A_3=-A_4=-\frac{Q}{2 \sqrt{2}} \psi^2\,, 
\label{eq:eigenvalue1} \\
A_5=A_6=-A_7=-A_8=-\frac{Q}{\sqrt{2}}(1- \psi)\,,
\end{eqnarray}
and
\begin{eqnarray}
&&8 \left(2 \pm \sqrt{2} \right)^{1/2} a_{1,3}=(4-\psi^2)(d_{-1
\uparrow} + d_{2\uparrow}) + [(4 \pm 2 \sqrt{2}) \psi +\nonumber\\
&&+(4 \pm
\sqrt{2}) \psi^2]
(d_{0 \uparrow}+d_{1 \uparrow})+(1 \pm \sqrt{2})(4-\psi^2)(d_{-1
\downarrow} - d_{2 \downarrow}) + \nonumber\\
&&+[\mp 2\sqrt{2}\psi-(2 \pm 3\sqrt{2}) \psi^2] (d_{0 \downarrow}
- d_{1 \downarrow})\,,\\
&&8 \left(2 \mp \sqrt{2} \right)^{1/2} a_{2,4}=\!-(4\!-\!\psi^2)(d_{-1
\uparrow}\! -\! d_{2\uparrow})\! +\! [(4 \mp 2 \sqrt{2}) \psi\! +\nonumber\\
&&+(4 \!\mp\!
\sqrt{2}) \psi^2]
(d_{0 \uparrow}\!-\!d_{1 \uparrow})\!+\!(-1\! \pm \sqrt{2})(4-\psi^2)(d_{-1
\downarrow} + d_{2 \downarrow}) + \nonumber\\
&&+[\pm 2\sqrt{2}\psi-(2 \mp 3\sqrt{2}) \psi^2] (d_{0 \downarrow}
+ d_{1 \downarrow})\,,\\
&&8 \left(2 \mp \sqrt{2} \right)^{1/2} a_{5,7}=\!-2\sqrt{2}
(\psi\!+\!\psi^2)(d_{-1 
\uparrow}\! + d_{2\uparrow})\!+\!(\sqrt{2}\! \mp\! 1)\!\times\!\!\!\!\!  
\nonumber\\
&&\times [4 \!\pm 2(\sqrt{2}\! \pm \!1)\psi\!  - \frac{3}{2}\psi^2]
(d_{0 \uparrow}\!+\!d_{1 \uparrow})\pm(4 \mp
2\sqrt{2})(\psi+\psi^2)\times  \nonumber\\ 
&&\times(d_{-1
\downarrow} - d_{2 \downarrow}) 
+[\pm 4 - 2 (\sqrt{2}\mp 1) \psi \mp \frac{3}{2}\psi^2] (d_{0
\downarrow} - d_{1 \downarrow})\,,\\
&&8 \left(2 \pm \sqrt{2} \right)^{1/2} a_{6,8}=2\sqrt{2}
(\psi\!+\!\psi^2)(d_{-1 
\uparrow}\! -\! d_{2\uparrow})\!+\!(\sqrt{2}\! \pm \!1)\!\times  \nonumber\\
&&\times [4 \!\mp\! 2(\sqrt{2}\! \mp\! 1)\psi\!  - \frac{3}{2}\psi^2]
(d_{0 \uparrow}\!-\!d_{1 \uparrow})\!\pm\!(4 \pm
2\sqrt{2})(\psi\!+\!\psi^2)\times  \nonumber\\ 
&&\times(d_{-1
\downarrow} + d_{2 \downarrow}) 
+[\mp 4 - 2 (\sqrt{2}\pm 1) \psi \pm \frac{3}{2}\psi^2] (d_{0
\downarrow} + d_{1 \downarrow})\,.
\label{eq:eigenvector8}
\end{eqnarray}
These expressions are then used to form the matrix elements $M_{ij}$
(see Eq. (\ref{eq:mij})), for example
\begin{eqnarray}
&&M_{11}=\frac{2-\sqrt{2}}{4Q}(4E_\uparrow^2-1)-\frac{2+\sqrt{2}}{4Q}(4E_\downarrow^2-1)+(2-\sqrt{2})\times
\nonumber \\
&&\times(1+3E_\uparrow-4E_\uparrow^3)I_\uparrow+
(2+\sqrt{2})(1-3E_\downarrow+4E_\uparrow^3)I_\downarrow+{\cal O}(\psi),\nonumber \\
&&I_\alpha=\frac{1}{8\pi Q}\int\frac{dk_y}{E_\alpha+\cos k_y- {\rm i}0}
\end{eqnarray}
[see Eq. (\ref{eq:Eupdown})]. Note that, owing to the symmetry
properties of the operators $a_i$,
the quantities $M_{ij}$ vanish unless both indexes $i$ and $j$ are
either odd or even. Hence the 8$\times$8 determinant on the r.\ h.\
s. of Eq. (\ref{eq:xigeneral}) reduces to a product of two 4$\times$4
determinants. After some algebra, one obtains expression
(\ref{eq:xifinal}), which has to be substituted into
Eqs. (\ref{eq:diagabrupt}) and (\ref{eq:sigmadiag}).

In the case of a vertical wall we choose the co-ordinate axes $r_1$
and $r_2$ along the lattice directions with the $r_2$ axis
perpendicular to the wall. After the Fourier transformation,
\begin{equation}
d_\alpha(r_1,r_2) = N^{-1/4}\sum_{k_1}{\rm e}^{\rm i k_1 r_1} d_\alpha
(k_1,r_2)\,,
\end{equation}
we find that the unperturbed Hamiltonian has the form 
\begin{equation}
\tilde{\cal H} = \sum_{k_1}\left({\cal H}_{k_1} - \cos k_1 \sum_{r_2}
d^\dagger_\alpha (k_1, r_2) d_\alpha(k_1,r_2) \right)\,, 
\label{eq:Ham4}
\end{equation}
and domain wall again results in a local perturbation, ${\cal H}_{k_1}
\rightarrow  {\cal H}_{k_1} + V_{k_1}$. This perturbation is still
illustrated by Fig. \ref{fig:abrupt1d}, although the intersite distance is now
equal to unity, rather than to $1/\sqrt{2}$. The operators ${\cal H}_{k_1}$
and $V_{k_1}$ have the same form as ${\cal H}_{k_x}$ and $V_{k_x}$
[see Eqs. (\ref{eq:Ham3}--\ref{eq:perturb})],
with the substitutions $Q \rightarrow 1$, $d_\alpha(k_x, y+1/\sqrt{2})
\rightarrow d_\alpha(k_1,y+1)$, $d_\alpha(k_x,i/\sqrt{2}) \rightarrow
d_\alpha (k_1,i)$.  Hence the
Eqs. (\ref{eq:eigenvalue1}--\ref{eq:eigenvector8}) with the value of $Q$ set 
to unity can be used to diagonalise the perturbation [see
Eq. (\ref{eq:diag})] in the case of a vertical wall as well.

We note that in the expressions for both ${\cal H}_{k_1}$
and $V_{k_1}$ in terms of operators $d_\alpha(k_1,r_2)$, the coefficients do
not depend on $k_1$. Therefore, the only
effect of the 
second term on the r.\ h.\ s. of Eq. (\ref{eq:Ham4}), regardless of
whether the domain wall is present, is to shift all of the energy
levels by $-\cos k_1$. 
Thus, Eqs. (\ref{eq:Omega0}--\ref{eq:trace1}) are now replaced by 
\begin{eqnarray}
\Omega &=&\int \frac{L_1 dk_1}{2 \pi} \int d \epsilon_2 \nu_{tot2}
(\epsilon_2) \varphi(\epsilon_2+\epsilon_1) 
\,, \nonumber\\
\frac{\delta \Omega}{L_1} &=& \int \frac{dk_1}{2 \pi} \int d
\epsilon_2 \delta \nu_2 (\epsilon_2) \varphi(\epsilon_2+\epsilon_1) =
\nonumber \\
&=& \int\frac{dk_1}{2
\pi} \int d 
\epsilon_2 \tilde{\xi}(\epsilon_2) f(\epsilon_2+\epsilon_1)\,, \\
\epsilon_{1,2} &=& - \cos k_{1,2}\,,\,\,\,\,\, \nu_{tot2}
(\epsilon_2)=L_2/(\pi \sqrt{1-\epsilon_2^2})\,.\nonumber
\end{eqnarray}
Here, $L_1$ and $L_2$ are the dimensions of the sample, and
$\tilde{\xi}(\epsilon)$ is the spectral shift function of the
corresponding 1D problem. It is evaluated as
$\pi \tilde{\xi}(\epsilon)=-{\rm Arg Det}(\delta_{ij}-\tilde{M}_{ij}A_j)$, 
with 
\begin{equation}
\tilde{M}_{ij}= \sum_{\alpha=\uparrow,\downarrow} \int \frac{dk_2}{2
\pi} \frac{\langle 0 |a_i |k_2^\alpha \rangle \langle k_2^\alpha |
a_j^\dagger | 0 \rangle}{E_\alpha + \cos k_2 - {\rm i}
0}
\end{equation}
and $E_\uparrow=E_\downarrow+J_H=\epsilon$.
Taking also into account that the states $|k_2^\alpha\rangle$ are defined in a
conventional way, $|k_2^\alpha\rangle= \sum_{r_2} {\rm exp}(-{\rm
i}k_2 r_2 )d^\dagger_\alpha (k_1,r_2) |0 \rangle  $
[cf. Eq. (\ref{eq:wavediag})], we conclude that the value of
$\tilde{M}_{ij}$ coincides with that of $M_{ij}$, Eq. (\ref{eq:mij}),
calculated at $Q=1$. Thus, $\tilde{\xi}(\epsilon)=\xi(\epsilon, Q=1)$ [see
Eq. (\ref{eq:xifinal})], and Eqs. (\ref{eq:vertabrupt}),
(\ref{eq:sigmavert}) follow.

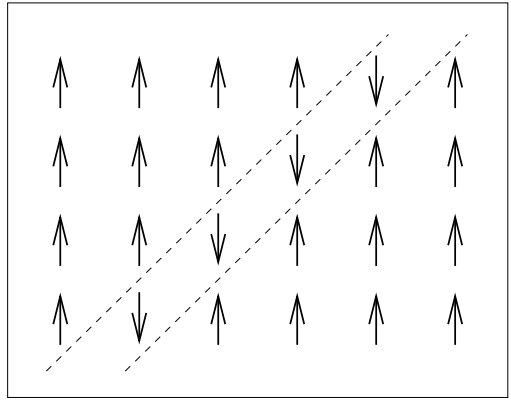
\begin{figure}
\caption{Local
perturbation used in the calculation of abrupt domain wall energies at
$J_H \rightarrow \infty$ (for the case of a diagonal wall).}
\label{fig:infjscheme}
\end{figure}

The somewhat cumbersome expressions
(\ref{eq:xifinal}--\ref{eq:sigmavert}) become much simpler in the case
of large Hund's rule coupling, $J_H \rightarrow \infty$. We then
find\cite{misprint} 
\begin{eqnarray}
&&S_d= \frac{1}{\sqrt{2}} \left\{ \frac{\sqrt{4-\mu^2}}{\pi}-\frac{|\mu|}{\pi}
{\rm arc
cos}\frac{|\mu|}{2}-4J+ \right. \nonumber \\
&& +E-\mu[x-\theta(\mu)] \Bigg\} + 2 \sqrt{2}\bigg\{2J+K+ \nonumber \\
&& \left.+\frac{4}{3
\pi^2}\left[\frac{\mu^2}{2}Y_1-\left(1+\frac{\mu^2}{4}\right) Y_2
\right]\right\} \psi^2\,,
\label{eq:diagabruptinfj}\\
&&S_v=\frac{\sqrt{2|\mu|-\mu^2}}{2\pi}+\frac{1-|\mu|}{2\pi}{\rm arc
cos}(|\mu|-1)-2J +E-\nonumber \\
&&-\mu[x-\theta(\mu)] +4\left\{J+K-\frac{4}{9 \pi^2} \left[ \left(\mu^2+ \frac{3}{8} \mu^4 \right)
Y_1+ \right. \right. \nonumber \\
&&\left. \left. +\left(1-\frac{11}{4}\mu^2 \right)Y_2 \right]\right\}
\psi^2 \,,
\label{eq:vertabruptinfj}\\
&&\sigma_d= - \frac{e}{\pi \sqrt{2}} {\rm arc cos} \frac{|\mu|}{2} {\rm
sgn} \mu- \frac{e}{\sqrt{2}}(x- \theta(\mu))\,, \\
&&\sigma_v= - \frac{e}{2 \pi} {\rm arc cos} (|\mu|-1) {\rm
sgn} \mu- e(x- \theta(\mu))\,,
\label{eq:sigmavertinfj}
\end{eqnarray}
where $Y_i$ and $E$ are given by
Eqs. (\ref{eq:elliptic}--\ref{eq:bandenergy}). When deriving the 
$\psi=0$ values of $S_d$ and $S_v$ above, it is convenient to use a
calculation scheme somewhat different from that used in the
finite-$J_H$ case. Namely, the local perturbation we consider now
(see Fig. \ref{fig:infjscheme}) corresponds to inverting the spins
along a 1D chain, not only shutting the carrier hopping but also 
introducing a single chain of an
antiferromagnetic phase. The latter circumstance can easily be accounted
for by subtracting the difference of thermodynamic potentials
between the antiferro- and ferromagnetic phases; the advantage of this
method lies in a very 
simple form of spectral shift function, $\xi(\epsilon,Q)=
(1/2){\rm sgn} \epsilon$, corresponding to the perturbation shown in
Fig. \ref{fig:infjscheme}.

Another potentially important case when the integration in
Eqs. (\ref{eq:diagabrupt}--\ref{eq:sigmavert}) can be carried out
analytically is that of small electron densities, $x \ll 1$. 
For any value of $J_H \gg x$, we obtain:
\begin{eqnarray}
&&S_d \approx \frac{4 \sqrt{\pi}}{3} x^{3/2} - \pi \sqrt{\frac{J_H+4}{2
J_H}} x^2 - 2 \sqrt{2}J + \nonumber \\
&&+\sqrt{2}\left\{\pi J_H \left(1-\sqrt{1+\frac{4}{J_H}} \right) x^2 +
4J+2K \right\} \psi^2 \,,
\label{eq:smallxdiag} \\
&&S_v\approx \frac{4 \sqrt{\pi}}{3} x^{3/2} - \pi \sqrt{\frac{J_H+2}{
J_H}} x^2 - 2 J + \nonumber \\
&&+4\left\{\pi J_H \left(1-\sqrt{1+\frac{2}{J_H}} \right) x^2 +
J+K \right\} \psi^2 \,.
\label{eq:smallxvert} 
\end{eqnarray}
It is instructive to note that  expansion of these
expressions in the case of $J_H \gg 1$ shows that the leading
$1/J_H$ correction amounts to a renormalisation of the superexchange constant,
$J \rightarrow J+ (\pi x^2)/(2 J_H)$. This is another illustration of an
effective antiferromagnetism being induced by a finite Hund's rule coupling,
as discussed in Sect. \ref{sec:bloch}.
We also see that at $J=0$ and to leading order in $x\ll 1$,
abrupt wall energy does not depend on 
orientation of the wall, $S_d^{(0)}=S_v^{(0)}$, which is due to  the
carrier dispersion law being isotropic at low densities.

Electric charges of unperturbed abrupt domain walls at $x \ll 1, J_H$
are given by
\begin{eqnarray}
\sigma_d\approx e \sqrt{\frac{x}{\pi}} - e x \sqrt{\frac{J_H+4}{2J_H}}
-ex^{3/2}\frac{\sqrt{\pi}}{24} \,,\\ 
\sigma_v\approx e \sqrt{\frac{x}{\pi}} - e x \sqrt{\frac{J_H+2}{J_H}}
+ex^{3/2}\frac{\sqrt{\pi}}{24} \,.
\label{eq:smallxsigmavert}
\end{eqnarray} 
Finally, we also quote a 3D result for a vertical abrupt
domain wall energy (per unit area) at $x\ll 1$ and $J_H \rightarrow \infty$:
\begin{equation}
S^{(3D)}_{v} \approx \frac{2^{1/3}3^{4/3}}{16}\pi^{5/3}x^{4/3} - 2
J\,,\,\,\,\,\psi=0\,.
\label{eq:smallx3dvert}
\end{equation}

\section{Stripe Walls and Screening}
\label{app:coulomb}

In this Appendix, we are concerned predominantly with investigation of
screening potentials and Coulomb energies of phase-separated states in a
two-dimensional conductor. Let the values of dielectric constants of
the media on both sides of conducting plane be $\epsilon_{d1}$ and
$\epsilon_{d2}$. The method of images enables one to evaluate the
potential of a point charge $q$ located at a distance $z$ from  a
plane separating the two dielectric media\cite{Volume8}. In the 
limit $z\rightarrow 0$, we find that
this potential at any point in space is given by $q/(\bar{\epsilon} s)$, where
$s$ is the distance from the charge and
$\bar{\epsilon}=(\epsilon_{d1}+\epsilon_{d2})/2$. We therefore
conclude that the electrostatic properties of this system are described by a
Poisson's equation of the form
\begin{equation}
\bar{\epsilon} \nabla^2 \varphi = - 4 \pi \rho(\vec{r})
\delta(z)\,. 
\label{eq:laplace}
\end{equation}
Here, $\vec{r}=\{x,y\}$ is the 2D radius-vector in the plane, $z$ axis is
perpendicular to the conductor, and $\nabla$ is the usual 3D
gradient. It is therefore only the effective 
dielectric constant, $\bar{\epsilon}$, that will affect the values of physical
quantities in this case [cf. Eq. (\ref{eq:2ddebye})].

We begin with evaluating the potential of a charged string within the
film. Assuming that the string coincides with the $x$ axis, we re-write 
Eq. (\ref{eq:laplace}) as
\begin{equation}
\bar{\epsilon}\nabla^2 \varphi(y,z) = [4 \pi e^2 \varphi(y,z) \nu_0 - 4 \pi \sigma
\delta(y)] \delta(z)\,.
\end{equation}
Here, $\sigma$ is the linear charge density of the string, and the
first term on the r.\ h.\ s. accounts for a screening charge arising
from the band energy shift by the electrostatic energy, $-e \varphi$.
This is a standard Thomas--Fermi treatment of screening, valid in the
long-wavelength limit. Upon one-dimensional Fourier transformation we
obtain 
\begin{equation}
\left(\frac{\partial^2}{\partial z^2}-k_y^2\right)\varphi(k_y,z) = \left(2
\kappa \varphi(k_y,z)- \frac{4 \pi \sigma}{\bar{\epsilon}} \right) \delta(z).\,\,\,\,
\label{eq:laplace3}
\end{equation}
Using the Green's function for Eq. (\ref{eq:laplace3})
(cf. Ref. \cite{Fetter}),
\begin{equation}
g(k_y,z)=\int_{-\infty}^\infty \frac{dk_z}{2 \pi} \frac{{\rm e}^{{\rm i}k_z
z}}{k_z^2+k_y^2} = \frac{1}{2|k_y|}{\rm e}^{-|k_y z|}\,,
\label{eq:green}
\end{equation}
we obtain
\begin{equation}
\varphi(k_y,z)=\frac{1}{|k_y|}{\rm e}^{-|k_y z|} \left(\frac{2 \pi
\sigma}{\bar{\epsilon}} - \kappa \varphi (k_y,0) \right). 
\label{eq:potential2d}
\end{equation}
Hence at $z=0$, $\varphi(k_y)= 2 \pi
\sigma/[\bar{\epsilon}(|k_y|+\kappa)]$, and
\begin{equation}
\varphi(y)=\frac{-2 \sigma}{\bar{\epsilon}} \left( \cos \kappa y \,{\rm
ci} \kappa y + \sin  \kappa y \,{\rm
si} \kappa y \right)\,
\label{eq:string}
\end{equation}
where ${\rm si}$ and ${\rm ci}$ are sine and cosine integrals. At
$\kappa y \gg 1$, Eq. (\ref{eq:string}) yields $\varphi(y) \approx 2
\sigma/(\bar{\epsilon} \kappa^2 y^2)$. Along with the $1/r^3$ decay of a
screened point charge potential\cite{Stern,Fetter}, this is in
contrast  with the well-known exponential behaviours found in 3D.

Let us now consider the potential of an antiferromagnetic stripe of
width $d \ll \kappa^{-1}$, centred around the $x$ axis. At $y \gg
d$, it is  given by Eq. (\ref{eq:string}) with $\sigma
=\rho_{AFM}d$, whereas at $y \ll \kappa^{-1}$ it should coincide
with the unscreened potential of the stripe, 
\begin{eqnarray}
&&\varphi(y)=-\frac{2 \rho_{AFM}}{\bar{\epsilon}} \left
[ (\frac{d}{2}-y) {\rm ln} 
|\frac{d}{2}-y| + \right. \nonumber \\
&&\left. +(\frac{d}{2}+y) {\rm ln}
|\frac{d}{2}+y|-d \right]+ const\,,
\label{eq:stringunscreened}
\end{eqnarray}
At $y \gg d$, the latter expression takes the familiar form,
\begin{equation}
\varphi(y)=-\frac{2 \sigma}{\bar{\epsilon}} {\rm ln} |y| + const\,.
\label{eq:stringtextbook}
\end{equation}
The two regions, $y \gg d$ and $y \ll \kappa^{-1}$, overlap, enabling
us to find the value of $const$ in
Eqs. (\ref{eq:stringunscreened}--\ref{eq:stringtextbook}), 
$-2 \rho_{AFM} d(C+{\rm ln} \kappa)/\bar{\epsilon}$ where $C \approx
0.577$ is the Euler's constant. Substituting Eq.(\ref{eq:stringunscreened})
into Eq. (\ref{eq:loneisland2}), we find the leading order expression for 
the Coulomb energy of the stripe [the last term in Eq. (\ref{eq:lonestripe})]. 
It is also easy to estimate the Coulomb energy of a Bloch wall,
\begin{equation}
-\frac{1}{\bar{\epsilon}} \int\int_{- \infty}^{\infty} \delta x(y)
 {\rm ln}{|\kappa (y-y^\prime)|} \delta x (y^\prime) dy d y^\prime\sim \frac{1}{\bar{\epsilon}}
\sigma_B^2 {\rm
ln} |\kappa l_B|
\label{eq:blochcoulomb}
\end{equation}
[see Eq. (\ref{eq:deltax})],
assuming that $\kappa l_B  \ll 1$.

We note that screening affects the value of potential at $2|y|< d$
even in the limit of $d \ll \kappa^{-1}$
because the unscreened potential, Eq. (\ref{eq:stringtextbook}),
diverges at $y \rightarrow \infty$. This is also the case for the
potential of a single antiferromagnetic layer of thickness $d$ in a
phase-separated sample in three dimensions, in which case we find, per
unit area,
\begin{equation}
\frac{1}{2} \int_{AFM} \rho \varphi d^3 r \approx
\frac{\pi d^2 \rho_{AFM}^2}{\kappa_{(3D)} \epsilon_d}\,,\,\,\,\,
\kappa_{(3D)}^2=\frac{4 \pi e^2
\nu_0}{\epsilon_d}\,.
\label{eq:3dlayer}
\end{equation}
In the case of a single antiferromagnetic disk in 2D, or an
antiferromagnetic sphere (ball) in 3D, the unscreened potential vanishes at
large distances and the leading-order (in $\kappa R$) term  in the
Coulomb energy does not depend on the screening radius. Indeed, for
 the 2D case  at sufficiently small distances $r \ll
\kappa^{-1}$, the exact screened potential of a point charge
$\pi R^2 \rho_{AFM}$ (found in Refs. \cite{Stern,Fetter}) is to leading
order given by $\pi R^2 \rho_{AFM}/(\bar{\epsilon} r)$. For $r \gg R$,
this clearly matches the unscreened potential of a charged disk.
Therefore screening does not affect the value of $\varphi$ within  the
disk, which enters Eq. (\ref{eq:loneisland2}). 
Using the Green's function procedure similar to
Eqs. (\ref{eq:laplace3}--\ref{eq:green}) above, we obtain for an
unscreened disk of radius $R$
\begin{equation}
\varphi(r)=\int\varphi(k)e^{{\rm i}\vec{k} \vec{r}} 
\frac{d^2k}{4\pi^2}\,,\,\,\,\varphi(k)=\frac{4 \pi^2 \rho_{AFM}
R}{\bar{\epsilon} k^2} J_1(kR)
\label{eq:loneisland2.1}
\end{equation} 
(where $r$ is the distance from the island centre, and $J_1$ is Bessel
function), and
\begin{equation}
\frac{1}{2}\int\frac{d^2 k}{4 \pi^2} \rho(k) \varphi(k) =
\frac{2 \pi^2 \rho_{AFM}R^2}{\bar{\epsilon}}\int_0^\infty
\left[J_1(kr)\right]^2 \frac{dk}{k^2}\,,
\label{eq:loneisland2.2} 
\end{equation}
leading to Eq. (\ref{eq:loneisland3}),
whereas for a 3D sphere of radius $R$ we readily find
\begin{equation}
\frac{1}{2} \int_{AFM} \rho \varphi d^3 r \approx \frac{16 \pi^2
\rho_{AFM}^2 R^5}{15 \epsilon_d}\,.
\label{eq:3dball}
\end{equation}
Eq. (\ref{eq:threshold}), derived in
Sect. \ref{sec:abrupt}, holds for a thin film. 
With the help of Eqs. (\ref{eq:3dlayer}) and (\ref{eq:3dball}), it is
easy to obtain a similar phase-separation threshold condition for the 3D
(bulk crystal) case,
\begin{equation}
\Omega_{FM}-\Omega_{AFM}> \Delta_0^{(3D)}=\left(\frac{3^5 \pi
W_{(3D)}^2 \rho_{AFM}^2}{5 \epsilon_d} \right)^{1/3}.
\label{eq:threshold3d}
\end{equation}
Here, $W_{(3D)}$ is the energy per unit area of a
ferromagnetic-antiferromagnetic boundary in three dimensions, which
can be approximated by half the value of the 3D abrupt wall energy
(cf. Eq.(\ref{eq:smallx3dvert})). The energy of stripe (layer) domain wall at
the phase-separation threshold in 3D is then given by
[cf. Eq. (\ref{eq:stripewall1})] 
\begin{equation}
S_s^{(3D)}=2W_{(3D)}\left(1-\frac{27}{20}\kappa_{(3D)}R_0^{(3D)}\right)\,,
\label{eq:stripewall13D}
\end{equation}
per unit area,
where 
\[R_0^{(3D)}=\left(\frac{15 \epsilon_d W_{(3D)}}{8 \pi \rho_{AFM}^2}
\right)^{1/3}\]
 is the radius of antiferromagnetic bubbles appearing
immediately above the threshold, Eq. (\ref{eq:threshold3d}).  

We now turn to the other regime of phase separation considered in
Sect. \ref{sec:stripe}. In this case, screening is negligible and our
estimates of Coulomb contributions to thermodynamic potential are
based on evaluating the electrostatic energy of a single unscreened
Wigner cell. In the case of circular antiferromagnetic islands
(``droplet phase''), the Fourier component of electric potential of a
Wigner cell is given by [cf. Eq. (\ref{eq:loneisland2.1})]
\begin{equation}
\varphi(k)=\frac{4 \pi^2 R^\prime}{\bar{\epsilon}k^2 R}\rho_{FM} \left[R
J_1(kR^\prime) - R^\prime J_1(kR)\right].
\end{equation}
Momentum integration [cf. Eq. (\ref{eq:loneisland2.2})] then yields
the expression for the 
Wigner cell energy $E_1$, given in the text above
Eq. (\ref{eq:omega1.2}), with 
\begin{equation}
\frac{A_1(\delta)}{1+\delta}=1+\sqrt{\frac{\delta}{1+\delta}}\left[1-\frac{3
\pi}{4}\,_2F_1\left(\frac{1}{2},-\frac{1}{2};2;\frac{\delta}{1+\delta}\right)
\right]\,,  
\label{eq:a1delta}
\end{equation}
where $\,_2F_1$ is the hypergeometric function.

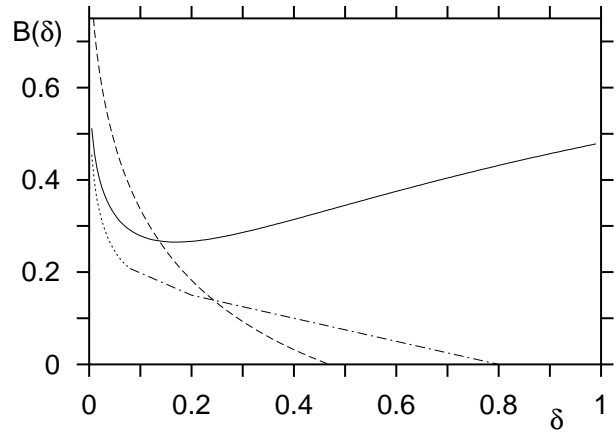
\begin{figure}
\caption{The ratio $B$ of the energy of a stripe wall to that
of an abrupt wall [see Eq. (\ref{eq:stripe2})] vs. the ratio $\delta$
of antiferro- and ferromagnetic areas of the sample: solid line,
droplet phase; dotted line, square-droplet phase at low $\delta$;
dashed-dotted line, possible behaviour for the square-droplet phase at
larger $\delta$; dashed line, 3D result of Eq. (\ref{eq:stripe2wall3d}).}
\label{fig:stripeplot}
\end{figure}

For the stripe phase, we find the electric potential $\varphi(y)$ of a single
``Wigner stripe'',
\begin{eqnarray}
\frac{\bar{\epsilon} \varphi(y)}{\rho_{AFM}}& =& 2y \delta{\rm ln}
\left|\frac{d^\prime + 2y}{d^\prime - 2y}\right|-2y (1+\delta){\rm ln}
\left|\frac{d + 2y}{d- 2y}\right|+ \nonumber \\
&&+d^\prime \delta {\rm ln}\left|
\frac{(d^\prime)^2-4 y^2}{d^2-4y^2}\right|\,,
\end{eqnarray}
where $y$ is the distance from the stripe centre. We note that at
$y \gg d^\prime$, the potential $\varphi(y)$ decays as $1/y^2$.
Evaluating the electrostatic energy per unit length,
$E_2=\int_{-d^\prime/2}^{d^\prime/2} \varphi(y) \rho(y) dy/2$, we obtain
the expression given in the text following Eq. (\ref{eq:omega2.1}), where
\begin{equation}
A_2(\delta){\rm ln} \delta=(1+\delta){\rm ln}  \frac{4
\delta(1+\delta)}{1+2 \delta}  -\frac{1+2\delta + 2 \delta^2}{2\delta}
{\rm ln}(1+2\delta)\,.
\label{eq:a2delta}
\end{equation}
It is also possible to evaluate the Coulomb energy of the stripe phase exactly,
taking into account the interaction between different ``Wigner
stripes''. Numerical calculation shows that this leads to an increase of
the
quantity $A_2$ by about $8\%$ at $\delta\rightarrow 1$, and by only
$3\%$ at $\delta=0.17$ [the latter corresponds to the minimum of $B(\delta)$ in
Fig. \ref{fig:stripeplot}], attesting to the relatively high
accuracy of the Wigner cell method for the stripe phase even in 2D. 

Coulomb energies of droplet and layered phase-separated states in 3D
were evaluated in Ref. \cite{Nagaev74}. The sum of boundary and
Coulomb contributions to the thermodynamic potential equals
\begin{eqnarray}
\tilde{\Omega}_1^{(3D)}(R)&=&\frac{3 \delta}{1+\delta} \frac{W_{(3D)}}{R}
+ \frac{4\pi}{5 \epsilon_d}R^2 \rho_{AFM}^2
A_1^{(3D)}(\delta)\delta\,,
\label{eq:droplet3D} \\
A_1^{(3D)}&=&1+\frac{3}{2}\delta-\frac{3}{2}\delta^{1/3}(1+\delta)^{2/3}
\end{eqnarray}
for the droplet phase, and
\begin{equation}
\tilde{\Omega}_2^{(3D)}(d)=\frac{2 \delta}{1+\delta} \frac{W_{(3D)}}{d}
+ \frac{\pi}{6 \epsilon_d} d^2 \rho_{AFM}^2
\end{equation}
for the layered phase.
Minimising expression (\ref{eq:droplet3D}) with respect to the radius
$R$ of spherical antiferromagnetic droplets, and then the 3D stripe wall
energy per unit area,
$S_s^{(3D)}=(\tilde{\Omega}_2^{(3D)}(d_s)-\tilde{\Omega}_1^{(3D)})d^\prime_s$ 
[where $d^\prime_s= d_s(1+\delta)/\delta$], with respect to the
antiferromagnetic layer thickness $d_s$, we find
\[d_s^{(3D)}=\frac{3^{5/6}\cdot\sqrt{2\delta}(A_1^{(3D)})^{1/6}}
{5^{1/6} \pi^{1/3}|\rho_{AFM}|^{2/3}} \left(\frac{W_{(3D)}
\epsilon_d}{1+\delta}\right)^{1/3}\,,\]
\begin{equation}
S^{(3D)}_s=2W_{(3D)}B_{(3D)}(\delta)\,,\,\,\,\,
B_{(3D)}(\delta)=1-3 \sqrt{\frac{6}{5}A_1^{(3D)} \delta}\,.
\label{eq:stripe2wall3d}
\end{equation}
The ratio $B_{(3D)}(\delta)$ of the energies of stripe and abrupt walls in 3D
is plotted in Fig. \ref{fig:stripeplot} (dashed line). We see that the
stripe wall energy vanishes already within the Wigner-cell method as
the value of $\delta$ approaches $\delta_c^{(3D)} \approx 0.47$.

\begin{references}
\bibitem{axis} $k_1$ and $k_2$, directed along the principal directions
of the inverse lattice, are not to be mixed with diagonally directed
and rescaled $k_x$ and $k_y$, introduced later in the article.
\bibitem{qualify} We stress that magnetic properties of these two
systems are by no means identical, which can be seen from
Eq.(\ref{eq:magnon}) yielding a Heisenberg (cosine) dispersion law
at $J_H \rightarrow \infty$, but not for any finite $J_H$ value.
\bibitem{blochcharge} The integration variable $\epsilon_x$
corresponds to $\cos (k_x/2)$. We also assume that $\mu$ is negative;
the $\mu>0$ case is treated similarly.
\bibitem{energy} Bloch wall energy, i.e. the change in the
thermodynamic potential associated with the Bloch wall, can also be
evaluated in a similar way, thus confirming the classical result 
(\ref{eq:bloch}).
\bibitem{omit} By using the Hamiltonian (\ref{eq:Ham}), which does not
include Coulomb interactions.
\bibitem{hundbound} This holds for both finite and infinite values of
Hund's rule coupling. In the case of
$J_H= \infty$, the carrier wave functions must vanish at the wall,
whereas at finite values of $J_H$, exponential tail of spin-up wave
function extends into the spin-down magnetic domain, and vice versa
(see Eq. (\ref{eq:hop})).
\bibitem{3d} In three dimensions, one finds that the energy per unit
area of partition is $\sim x^{4/3}$ [see Eq. (\ref{eq:smallx3dvert})].
\bibitem{1d} It is because of 1D character of the perturbation that
one is able to introduce spectral shift function in this way;
Eq. (\ref{eq:trace1}) generalizes the standard Lifshits--Krein trace
formula \cite{iml}.
\bibitem{unstabletyp} At these values of $J$, the uniform
ferromagnetic state is typically unstable with respect to phase separation.
\bibitem{intermediate} We emphasize that this statement applies to the
intermediate doping range, $x,1-x \sim 1$.
\bibitem{J0} Adding a non-zero superexchange, $J>0$, would (i) result
in the diagonal (rather than vertical) abrupt wall being preferred at
low $x$ (ii) give rise to a phase-separation instability within a
certain low-$x$ region.
\bibitem{plane} Such a wave function does not necessarily have to be
a Bloch wave. In a purely 1D case, this would be a cosine wave, and it
is also important that its wavevector will have to be related to the
energy of the localised state by the usual tight-binding dispersion
relation. We note that the choice of the basic set of electron 
energy eigenstates
used in the Lifshits--Krein trace formalism [cf. Eqs. (\ref{eq:mij}) and
(\ref{eq:wavediag})] is, in principle, arbitrary.
\bibitem{singleband} We consider the single-band case [cf. Eq. (\ref{eq:Ham})],
when the carriers originate solely from doping, and not from band overlap.
\bibitem{zoo} Some possible examples of antiferromagnetic phases are
discussed in Ref. \cite{jap02}.
\bibitem{2Dscreen} This requirement is perhaps somewhat relaxed in the case
of a thin film, due to the power-law (rather than exponential -- see Appendix 
\ref{app:coulomb}) decay of screened Coulomb potential.
\bibitem{omitcharge} Strictly speaking, there is also a surface charge
(and hence the Coulomb energy) associated with the interphase boundary
itself, which in this respect is similar to abrupt domain walls considered in
Sect. \ref{sec:abrupt}. Throughout Sect. \ref{sec:stripe} we omit 
these electrostatic corrections to the energies of abrupt and Bloch domain
walls and interphase boundaries. They are expected to be much smaller than
the electrostatic energies of large charged islands considered here.
\bibitem{smallcorr} The numerical smallness of $W$ (which can be estimated as
half the energy of abrupt domain wall, see in the text below)
and the fact that the actual values of $\bar{\epsilon}$ are large result
in the value of $\Delta_0$ [see Eq. (\ref{eq:threshold})] being small, and more
generally, in $|\Omega_{FM}-\Omega_{AFM}| \ll |\Omega_{FM,AFM}|$. It is 
therefore consistent to assume $\Omega_{FM}=\Omega_{AFM}$ when evaluating
abrupt wall energies and $W$.
\bibitem{abrupt} We recall that the interphase boundaries are expected
to be abrupt \cite{jap02}.
\bibitem{point} This holds within the Wigner-cell approximation for
the stripe phase. Within a more rigorous treatment, $S_s$ will vanish at a
certain point near $\delta_c$.
\bibitem{thick3D} Although under certain conditions  stripe wall
formation may be possible in the opposite case of a thick film or bulk
crystal (see below and Appendix \ref{app:coulomb}), there is no reason
to expect that these conditions are satisfied for thicker samples
studied in Ref. \cite{Li}.
\end{references}
\end{document}